# Magnon-bipolar carrier drag thermopower in antiferromagnetic/ferromagnetic semiconductors: theoretical formulation and experimental evidence

Md Mobarak Hossain Polash[1,2] and Daryoosh Vashaee[1,2]

[1]Department of Materials Science and Engineering, North Carolina State University, Raleigh, NC 27606
[2]Department of Electrical and Computer Engineering, North Carolina State University, Raleigh, NC 27606

## Abstract

Quantized spin-wave known as magnon, a bosonic quasiparticle, can drag electrons or holes via s-d exchange interaction and boost the thermopower over the conventional diffusive thermopower. P-type magnon-drag thermopower has been observed in both ferromagnetic and antiferromagnetic metallic and degenerate semiconductors. However, it has been less reported for intrinsic or n-type magnetic semiconductors; therefore, the impact of magnon-bipolar carrier drag on thermopower has remained unexplored. Here, a theoretical model for magnon-bipolar carrier drag thermopower is derived based on the magnon-carrier interaction lifetimes. The model predicts that the bipolar carrier drag thermopower becomes independent of both the carrier and magnon relaxation times. A proof of concept experiment is presented that confirms this prediction. We also report the observation of magnon-carrier drag thermopower in n-type and intrinsic ferromagnetic semiconductors experimentally. The p-type antiferromagnetic MnTe is doped with different amounts of Cr to produce non-degenerate and n-type semiconductors of various carrier concentrations. Cr dopants have a donor nature and create ferromagnetic-antiferromagnetic clusters due to the $Cr^{3+}$ oxidation state. Heat capacity measurements confirm the presence of magnons in Cr-doped MnTe. It is shown that the magnon-drag thermopower is significantly reduced for 3%-5% Cr-doped samples due to bipolar drag effects and becomes negative for 14% and 20% Cr-doped MnTe due to dominant magnon-electron drag thermopower.

## Introduction

Coupling between spins or spin-wave with the charge carrier in the presence of phonon can give rise to some exciting entropy transport phenomena like magnon-electron drag [1,2], paramagnon-electron drag [3,4], spin-entropy [5,6], spin-fluctuation [7,8], etc., which offer routes to design high-performance spin-based thermoelectric materials. The spin degree of freedom can exist in different forms, the spin of electrons (conduction or valence electrons), spin of the lattice ions, and the spin waves (magnons) or spin-wave packets (paramagnons) resulted from the precession of the spin of the lattice ions. The coherent fluctuation of the long-range ordered localized spins form magnons, the bosonic quasi-particles of the quantized spin-wave, which can interact with the charge carriers in magnetic materials like ferromagnets (FM) or antiferromagnets (AFM) and lead to a magnon-enhanced thermopower [9]. The improvement of the thermopower due to the magnon electron drag near the magnetic transition temperature has been reported in several long-range ordered systems [1,2,4,10,12]. Recently, a significant spin contribution to thermopower in MnTe far above the Néel temperature was reported, which was associated with the paramagnon-drag effect [3,4,13,14]. Like magnon and paramagnon drag, spin entropy and spin fluctuation in a system has also led to significant enhancement of the thermopower [5-8].

It should be noted that these effects may be fundamentally related. These advantageous spin-caloritronic effects boost the investigations on magnetic and paramagnetic semiconductors for the development of high-performance thermoelectric materials [10-14]. To accelerate the search for better magnetic or paramagnetic semiconductors for thermoelectric applications, one should know the general thermoelectric performance trends in different types of material systems, such as intrinsic and extrinsic semiconductors. Till now, magnon-drag thermopower has been reported for mostly p-type semiconductors and rarely reported for n-type and intrinsic semiconductors. Therefore, the effects of magnon-drag in bipolar conduction systems remains unknown. In this work, we explored the impact of magnon-bipolar carrier drag on thermopower both theoretically and experimentally. We found that thermopower is significantly reduced due to the presence of magnon-bipolar carrier drag. We demonstrated the thermopower trends of Cr-doped MnTe as a case study to provide experimental support for the theoretical findings.

Among magnetic semiconductors, manganese telluride (MnTe), with NiAs hexagonal crystal structure, has been widely studied as an antiferromagnetic semiconductor doped with different materials [4,13-16]. In particular, p-type Li doped MnTe with the Néel temperature $T_N \sim 307K$ has shown a significant zT enhancement due to the spin-caloritronic effects [4,13,14]. Magnetic dopants from transition metal (TM) elements can further modify the electronic and magnetic properties of MnTe [17-19]. However, their effects on thermoelectric properties have not been studied much in the literature [18]. Among the TM dopants, Cr-doped MnTe is mostly considered due to its high solubility into MnTe [17-20]. MnTe and CrTe are both isostructural to NiAs system with about similar c/a ratio [17-19]. CrTe is a ferromagnetic semiconductor with a Curie temperature of 342 K [18,19]. CrTe and Cr-rich MnTe have metallic conduction with a small negative Seebeck coefficient [18]. Cr-doping can induce ferromagnetic ordering into the antiferromagnetic MnTe [21], which is also observed in some other Cr-doped materials [22]. In the $Mn_{1-x}Cr_xTe$ system, Mn-Mn and Mn-Cr couplings favor antiferromagnetic behavior, while Cr-Cr coupling favors ferromagnetic behavior [19,23]. Both bond-length and bond angles also determine the nature of the couplings [20]. With sufficiently high Cr concentrations, Cr-Cr ferromagnetic interactions are favored by hole-mediated exchange interaction [19,24]. Moreover, the common oxidation state of $Cr^{+3}$ can introduce n-type carriers into p-type MnTe [25], which can provide the bipolar conduction into MnTe system. Therefore, Cr is the right candidate to demonstrate the theoretically predicted magnon-bipolar carrier drag effects on thermoelectric performance. However, Cr-Cr ferromagnetic interaction can be a roadblock to observe the magnon-bipolar drag effect due to the suppression of the antiferromagnetic magnons of MnTe. For the detailed investigation, x% Cr-doped MnTe samples were synthesized with different x values ranging from 3 to 20. All samples were characterized versus their electrical, thermal, optical, and magnetic properties along with their thermoelectric properties. A detailed discussion is given to highlight the effects of Cr doping in the spin-mediated transport properties of MnTe, and a comparison is drawn for the thermoelectric performances of x% Cr-doped and undoped MnTe systems. In particular, the bipolar magnon drag is demonstrated for the first time and discussed regarding this material system.

**Formulation of the magnon-bipolar carrier drag thermopower in intrinsic semiconductors**

Bipolar carrier transport in semiconductors can be obtained by tuning the Fermi level with external stimuli like the electric field, magnetic field, heat, or doping [26,27]. Bipolar effect on transport properties is generally more significant in narrow bandgap and intrinsic semiconductors [28]. Bipolar conduction may exist at low temperature in intrinsic semiconductors as well as at high temperature due to the thermal carrier excitation [28]. Generally, electrical conductivity and thermal conductivity are enhanced due to the bipolar conduction, while the thermopower is reduced. It is worth noting that enhancement in electrical conductivity and thermal conductivity due to bipolar conduction happens for those systems where the intrinsic condition is attained by tuning the Fermi level without doping. In the doped samples, depending on the type of dopant, the electrical conductivity may increase by adding more carriers or decrease by compensating the carriers of the opposite charge. Moreover, doping can reduce the carriers' mobility by introducing ionized impurity scattering. On the other hand, thermopower always suffers from the bipolar effect due to the opposite charge of the bipolar carriers and due to the diffusion of both carriers in the same direction.

Different types of bipolar transport properties are illustrated in Figure 1. All the bipolar effects on transport properties except magnon-bipolar carrier drag are well known and have been theoretically and experimentally studied for different material systems [28,29]. In this work, we have formulated the magnon-bipolar carrier drag thermopower for the intrinsic magnetic semiconductors. As shown in the figure, the magnon-carrier drag nature resembles the phonon-carrier drag. Unlike the carrier diffusion due to the thermal gradient, in magnon-carrier drag, carriers are dragged by the magnon flux, which is caused by the thermal fluctuation of local magnetic moments in a magnetically ordered system. Both electrons and holes can be coupled with the magnon flux by s-d or p-d exchange interaction and can be dragged by magnons in the same direction as the magnon flux, which always follows the heat flow. The interaction between magnons and carriers is determined by the magnon-carrier scattering mechanisms, namely magnon-magnon scattering, magnon-on-electron scattering, and electron-on-magnon scattering. Magnon-magnon scattering includes all the magnon related scattering, magnon-on-electron scattering considers only the scattering of magnons by electrons, and electron-on-magnon scattering only considers the scattering of electrons by magnons. Magnon-bipolar carrier drag is a mixed effect, as the total thermopower has contributions from both diffusion and drag, where the scattering mechanisms related to the magnon-carrier drag can affect both of them. In this work, we are only considering the magnon-drag effect. Hence, here the drag thermopower in the formulation is always referring to the magnon-drag thermopower.

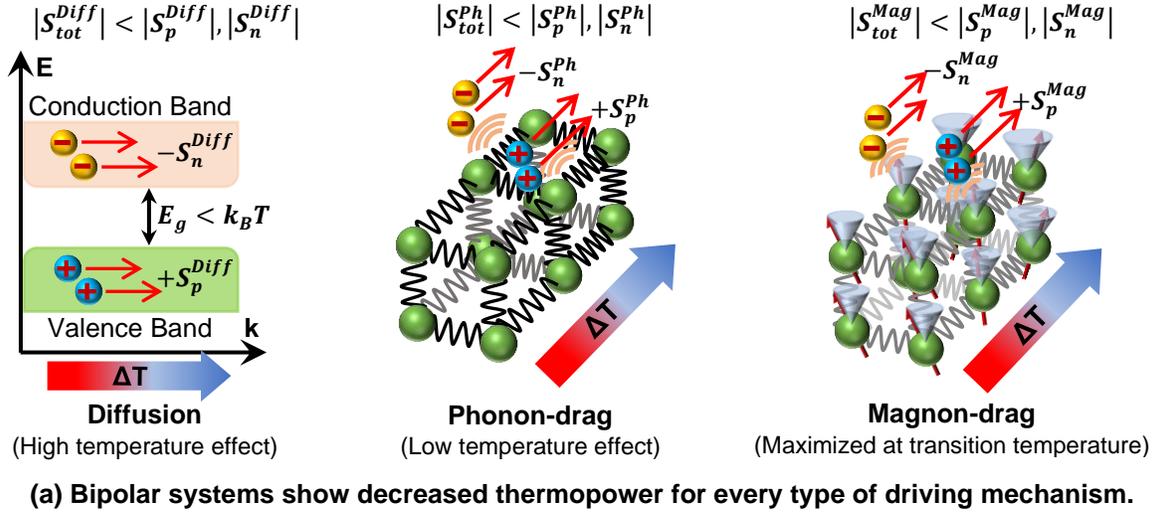

**(a) Bipolar systems show decreased thermopower for every type of driving mechanism.**

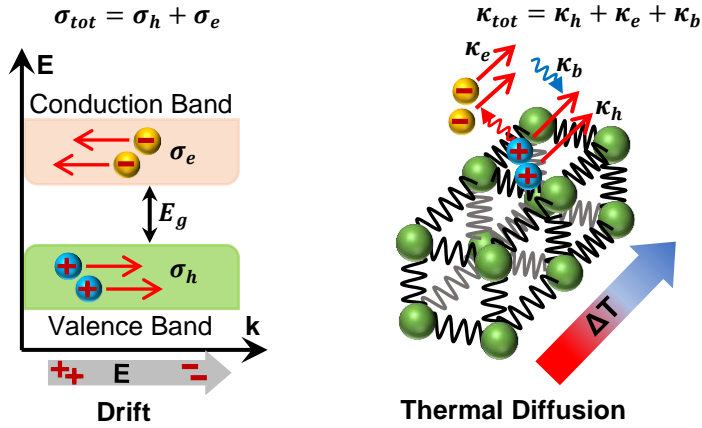

**(b) Both electrical and thermal conductivities increase by bipolar effects**

Figure 1: Impact of bipolar effects due to diffusion, phonon-drag and magnon-drag on the thermopower, electrical conductivity and thermal conductivity: Thermopower reduces, while electrical and thermal conductivity increase.

From the fundamental theory of diffusion and drag thermopower in intrinsic semiconductors, we can write the following expressions with modifications for the thermopowers due to diffusion and magnon-carrier drag [9,28,29]:

$$S_n^{diff} = -\frac{k_B}{e}\left[\left(\ln\frac{N_c}{n}\right) + \left(r + \frac{5}{2}\right)\right] \quad (1)$$

$$S_n^{drag} = -\frac{m_n^* c^2}{eT}\frac{\tau_m}{\tau_{em}}\frac{\tau_{me}}{(\tau_m + \tau_{me})} \quad (2)$$

Here, *n* is the carrier concentration, *r* is the electronic scattering parameter, *m\** is the effective carrier mass, *c* is the magnon group velocity, *T* is temperature, $\tau_m$ is the magnon lifetime due to all scatterings but charge carriers (magnon-magnon scattering), $\tau_{em}$ is the charge carrier relaxation time due to scattering by magnons (carrier-on-magnon scattering), and $\tau_{me}$ is the magnon relaxation time due to scattering by charge carriers (magnon-on-carrier scattering). $\eta = \ln\frac{N_c}{n} = \frac{E_c - E_F}{k_B T}$ is the reduced chemical potential energy of the carriers, while $\left(r + \frac{5}{2}\right)$ is considered as the reduced kinetic energy of the carriers. Similar expressions can be written for holes with a positive sign. In bipolar semiconductors, thermopower has contributions from both electrons and holes, and can be written as,

$$S = \frac{\left(S_n^{diff} + S_n^{drag}\right)\sigma_n + \left(S_p^{diff} + S_p^{drag}\right)\sigma_p}{\sigma_n + \sigma_p} = \frac{\Sigma^{diff} + \Sigma^{drag}}{\sigma_n + \sigma_p} \qquad (3)$$

$n$ and $p$ subscripts refer to the electrons and holes, respectively. The electrical conductivity, $\sigma$, has contributions from both the diffusion and magnon-carrier drag, i.e.:

$$\sigma_n = \frac{e^2 n}{m_n^*}\left(\frac{1}{\tau_{e0}} + \frac{1}{\tau_{em}}\right)^{-1}\left(1 + \frac{\tau_m}{\tau_{me}}\right) \approx \frac{e^2 n}{m_n^*}\tau_{em}\frac{\tau_m}{\tau_{me}} \qquad (4)$$

Where, $\tau_0$ is the carrier relaxation time due to all scatterings, except magnons. The 2nd term with the relaxation times ratio is due to the magnon drag second-order effect, as discussed for the thermopower. The inclusion of this term is to consider the momentum returned by the magnon system to electrons, which enhances the electron mobility. The expression of the electrical conductivity is simplified further with the assumptions that $\tau_m \gg \tau_{me}$ and $\tau_{e0} \gg \tau_{em}$ at $T \leq T_c$. Similar expressions are valid for holes.

Inserting the corresponding terms into the expression for $\Sigma^{diff}$ and $\Sigma^{drag}$:

$$\Sigma^{diff} = \frac{k_B}{e}\left[\left(ln\frac{N_v}{p}\right) + \left(r + \frac{5}{2}\right)\right]\frac{e^2 p}{m_h^*}\tau_{hm}\left(1 + \frac{\tau_m}{\tau_{mh}}\right)$$
$$- \frac{k_B}{e}\left[\left(ln\frac{N_c}{n}\right) + \left(r + \frac{5}{2}\right)\right]\frac{e^2 n}{m_e^*}\tau_{em}\left(1 + \frac{\tau_m}{\tau_{me}}\right) \qquad (5)$$

$$\Sigma^{drag} = \frac{m_h^* c^2}{eT}\frac{\tau_m}{\tau_{hm}}\frac{\tau_{mh}}{(\tau_m + \tau_{mh})}\frac{e^2 p}{m_h^*}\tau_{hm}\left(1 + \frac{\tau_m}{\tau_{mh}}\right)$$
$$- \frac{m_e^* c^2}{eT}\frac{\tau_m}{\tau_{em}}\frac{\tau_{me}}{(\tau_m + \tau_{me})}\frac{e^2 n}{m_e^*}\tau_{em}\left(1 + \frac{\tau_m}{\tau_{me}}\right) \qquad (6)$$

Considering, $\tau_m \gg \tau_{me}$, $\Sigma^{diff}$ can be simplified to:

$$\Sigma^{diff} = ek_B\tau_m\left(\left[\left(ln\frac{N_v}{p}\right) + \left(r + \frac{5}{2}\right)\right]\frac{p}{m_h^*}\frac{\tau_{hm}}{\tau_{mh}} - \left[\left(ln\frac{N_c}{n}\right) + -\left(r + \frac{5}{2}\right)\right]\frac{n}{m_e^*}\frac{\tau_{em}}{\tau_{me}}\right) \qquad (7)$$

For non-degenerate semiconductors, $\frac{\tau_{em}}{\tau_{me}} = 2\frac{m_e^* c^2}{k_B T} = K m_e^*$. Applying this relation for both electrons and holes, we can write,

$$\Sigma^{diff} = 2e\frac{c^2}{T}\tau_m\left(p\left(ln\frac{N_v}{p}\right) - n\left(ln\frac{N_c}{n}\right) + \left(r + \frac{5}{2}\right)(p - n)\right) \qquad (8)$$

Applying a similar condition of $\tau_m \gg \tau_{me}$ for the drag term, we will get:

$$\Sigma^{drag} = \frac{ec^2}{T} p\tau_m - \frac{ec^2}{T} n\tau_m = \frac{ec^2}{T} \tau_m(p-n) \qquad (9)$$

$\sigma_n + \sigma_p$ can also be simplified with the consideration of $\tau_{e0} \gg \tau_{em}$,

$$\sigma_n + \sigma_p \cong e\tau_m \left( \frac{ep}{m_h^*} \frac{\tau_{hm}}{\tau_{mh}} + \frac{en}{m_e^*} \frac{\tau_{em}}{\tau_{me}} \right) \qquad (10)$$

Inserting (8), (9) and (10) into (3), we can get the total thermopower expression for non-degenerate semiconductors,

$$S = \frac{k_B}{e(p+n)} \left( p\ln\frac{N_v}{p} - n\ln\frac{N_c}{n} + (p-n)(r+3) \right) \qquad (11)$$

This is the expression for the total thermopower of an intrinsic magnetic semiconductor. Two important conclusions can be drawn from the above thermopower relation: (i) the magnon-bipolar carrier drag effect significantly reduces the total thermopower compared to the thermopower of extrinsic magnetic semiconductor, and (ii) the total thermopower of intrinsic magnetic semiconductor is independent of the carriers and magnon scattering times. This indicates that the magnon-bipolar carrier drag effect in any magnetic semiconductor depends mainly on the carrier concentration, the effective masses, and the scattering exponent. The scattering exponent depends on the type of scattering mechanisms present in the system, and it can be temperature-dependent. The relaxation time, however, which is also temperature-dependent, depends on many parameters besides the scattering exponent, such as the number of scattering centers, the effective masses, and the carrier energy. Therefore, the scattering exponent and the relaxation time can have different impacts on the total thermopower, including the magnon-drag thermopower. According to the total thermopower expression, magnon-bipolar carrier drag reduces the total thermopower, which can come from the reduction of both diffusion and drag thermopower. Equating 11 to zero, one can find the conditions under which the total thermopower becomes zero due to magnon-bipolar carrier drag. Such a situation depends only on the mentioned parameters, i.e., the carrier concentration, the effective masses, and the scattering exponent irrespective of the scattering times present in the system. It should be noted that different magnon and electron related scatterings can happen in magnetic semiconductors like magnon-magnon, electron-electron, magnon-on-electron, and electron-on-magnon scatterings, but none are entering in equ. 11.

**Experimental Evidence for Magnon-bipolar Carrier Drag**

We have studied several x% Cr-doped MnTe samples (x = 3, 4, 5, 14, and 20) to demonstrate the magnon-bipolar carrier drag effect on thermopower experimentally. Details on material synthesis and characterization are discussed in the appendix. To investigate the magnon-bipolar carrier drag effect on thermopower, we will check for the existence of magnons in the system, determine the intrinsic carrier conduction nature, and analyze the thermoelectric transport properties. We will use the heat capacity data to observe the presence of magnons in the system, and the magnetic susceptibility data to understand the magnetic nature of the samples and to check for any magnetic phase impurities. We present the temperature and field-dependent magnetic susceptibility data for doped and undoped MnTe to determine the magnetic nature as well as to probe probable magnetic impurity phases. The heat capacity data for the synthesized samples are discussed in the following subsection to demonstrate the presence of magnons and other magnetic contributions if they exist. Hall carrier concentration is measured and reviewed to support the bipolar conduction type of the samples. Then, the thermoelectric transport data is presented to demonstrate the impact of the magnon-bipolar carrier drag effect on the thermopower of the synthesized samples.

MnTe is a p-type antiferromagnetic semiconductor with the Néel temperature of $T_N \sim 307K$ and a direct optical bandgap of around 1.25eV (see appendix). With the addition of Cr as a dopant into MnTe, n-type carriers can be introduced by substituting the host $Mn^{2+}$ ions by $Cr^{3+}$ ions, the common oxidation state of the Cr. Moreover, Cr ions can introduce the ferromagnetic phase into an antiferromagnetic MnTe host, which can be attributed to three probable reasons: (i) clusters of ferromagnetic/antiferromagnetic phases, (ii) a canted ferromagnetic structure, and (iii) a hole mediated exchange interaction. Due to the introduction of n-type carriers in p-type MnTe, Cr-doped MnTe is an ideal material system to achieve intrinsic magnetic semiconductors. As mentioned earlier, the bipolar nature in Cr-doped MnTe does not enhance the total carrier concentration, as $Cr^{3+}$ ions are mainly compensating the carrier contribution from $Mn^{2+}$ ions.

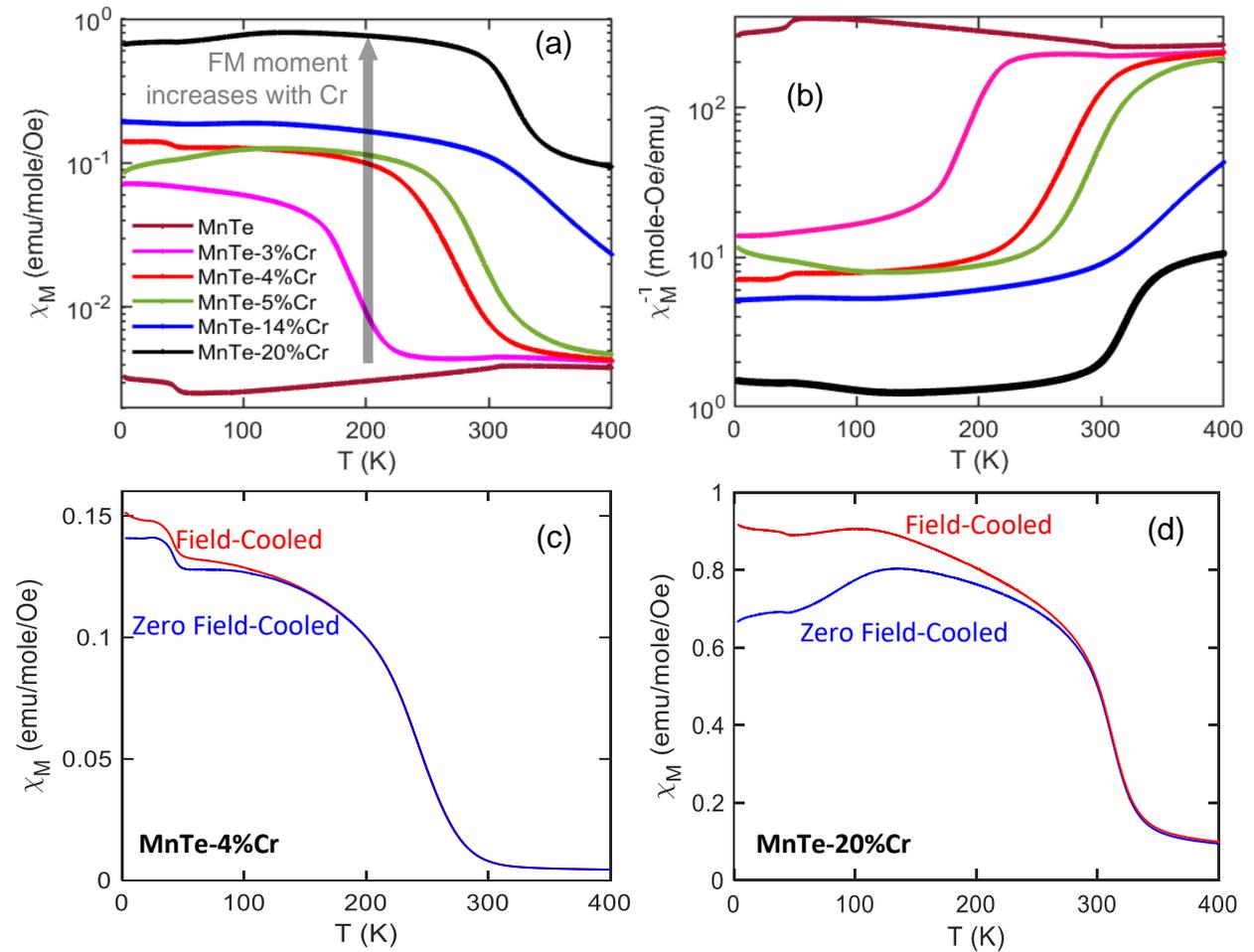

Figure 2: (a) and (b): Temperature dependent zero-field cooled magnetic molar susceptibility and its inverse, tespectively, for MnTe-x%Cr samples. (c) and (d): Zero-field and field cooled temperature dependent magnetic susceptibility of MnTe-4%Cr and MnTe-20%Cr, respectively. Susceptibility measurement is performed at 1000 Oe.

## Magnetic Nature of Cr-doped MnTe Samples

Figure 2(a) and (b) show the zero-field cooled (ZFC) magnetic molar susceptibility and its inverse for the different samples as a function of temperature, respectively. Figure 2(c) and (d) demonstrates temperature-dependent zero-field cooled (ZFC) and field-cooled (FC) magnetic molar susceptibility for MnTe-4%Cr and MnTe-20%Cr. The FC measurements were performed under the 12 T magnetic field. According to the magnetic susceptibility plots, MnTe is anti-ferromagnetic with a Néel temperature of approximately ~307K whereas MnTe-3%Cr, MnTe-4%Cr, MnTe-5%Cr, MnTe-14%Cr, and MnTe-20%Cr are ferromagnetic with a Curie temperature of around 174K, 259K, 278K, 337K, and 295K, respectively. The variations in the inverse magnetic susceptibility of MnTe-3%Cr indicate the co-existence of FM and AFM clusters. Here, Curie temperatures are calculated from the inverse magnetic susceptibility using Curie-Weiss law, $\chi = C/(T - T_c)$, where $\chi$ is the magnetic susceptibility and, $C$ is the Curie constant. A comparison table for Curie temperatures is given in the appendix with respect to the previously reported values [18,19,24]. The difference in transition temperatures with the values reported in the literature may be attributed to the presence of the small amount of a magnetic impurity phase as seen in the XRD data (see appendix), the co-existence of ferromagnetic and antiferromagnetic clusters [20], the calibration tolerance of the temperature sensors on either equipment or a combination of these effects. The estimated parameters of the Curie-Weiss law for MnTe-x%Cr samples are summarized in Table 1.

Table 1: Parameters of Curie-Weiss Law for MnTe-x%Cr samples

| Sample | MnTe | MnTe-3%Cr | MnTe-4%Cr | MnTe-5%Cr | MnTe-14%Cr | MnTe-20%Cr |
|---|---|---|---|---|---|---|
| C (emu/K/Mole) | 4.2 | 0.24 | 0.33 | 0.387 | 0.57 | 6.03 |
| $T_c/\theta$ (K) | 305/583 | 174 | 259 | 278 | 337 | 295 |
| $\mu_{eff}$ ($\mu_B$) | 5.79 | 1.38 | 1.62 | 1.76 | 2.14 | 6.9 |
| S | 2.44 | 0.35 | 0.45 | 0.51 | 0.68 | 3 |
| Hc (T) at 3K | -- | 0.17 | 0.11 | 0.34 at 3.5K | 0.24 | 0.05 |
| Br(emu/mole) at 3K | -- | 23.98 | 123.48 | 153.02 at 3.5K | 609.16 | 625 |

The magnetic moment of Cr-doped MnTe samples is increased with the increase of Cr doping, which can be coming from the formation of FM Cr-Cr exchange interaction, FM-AFM clusters, or the weak AFM Mn-Mn exchange interaction due to the low-spin (LS) state of $Mn^{2+}$ ions caused by the Cr-induced high crystal field (see Figure 3).

Cr-doping can create a FM impurity magnetic phase domain into the AFM MnTe (see Figure 3). CrTe magnetic phase impurity is also FM. If such impurity domains are small enough, they may have a superparamagnetic character and impact the overall magnetic properties of the samples. To investigate potential superparamagnetism existing in the Cr-doped MnTe samples, we measured the FC magnetic moments for both low-doped (4%Cr) and high-doped (20%Cr) samples. Considering the data shown in Figure 2 and Figure 4, one can see that FC and ZFC magnetic susceptibilities do not display any considerable change in the temperature-dependent magnetic nature of both 4%Cr and 20%Cr-doped MnTe samples. The field-dependent saturation magnetization does not show any feature of superparamagnetism for the Cr-doped MnTe either.

Therefore, it is unlikely that potential FM impurities are adding any significant superparamagnetic contribution to the overall magnetic nature of the Cr-doped MnTe.

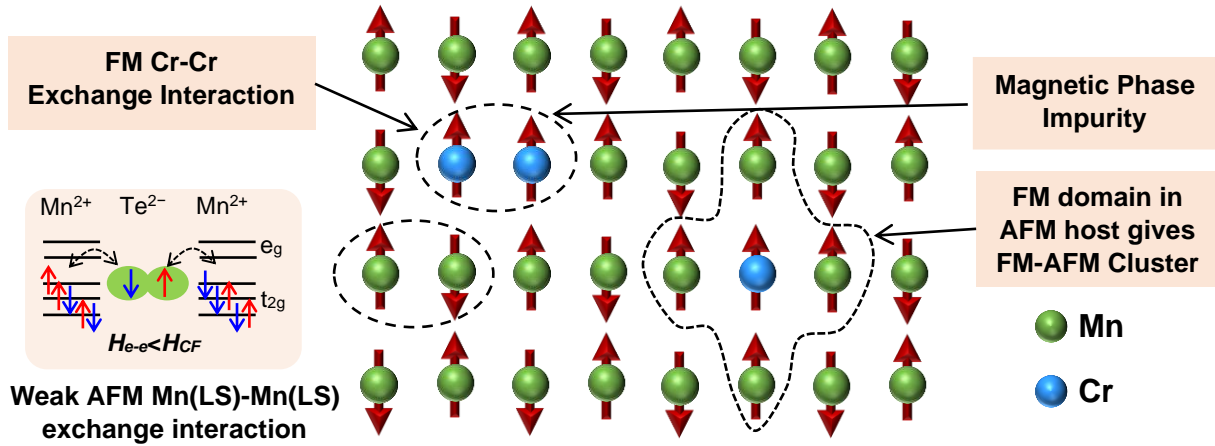

Figure 3: Illustration of possible reasons for variations in the magnetic moment of Cr-doped MnTe samples.

According to the table, Curie constant, Curie temperature, effective magnetic moment, and spin number of MnTe-x%Cr increases with $x$ $(x\neq0)$. MnTe and MnTe-20%Cr show different trends. In MnTe, $Mn^{2+}$ ions have a high spin state with an ideal spin number of 2.5 (note that zero-field splitting causes a lower spin number than ideal). Low spin numbers of MnTe-x%Cr, except for x=20, can be arisen from the induced low spin state of $Mn^{2+}$ due to the excess crystal field created by $Cr^{3+}$ ions. This further indicates that Cr has substituted Mn in the lattice. Anomalies observed in Curie-Weiss law for MnTe-20%Cr may be caused by the formation of the clusters into MnTe, complex magnetic dipole-dipole interaction between Mn-Mn or Mn-Cr ions, or the canted FM/AFM structures. Moreover, the existence of short-range or mid-range Cr-Cr ferromagnetic correlation in MnTe-20%Cr above the transition temperature (observed in Figure 4) may also arise the anomaly. The coercive field ($H_c$) and residual magnetic moment ($B_r$) are determined from Figure 4, which demonstrates the magnetic moment versus magnetic field (M-H) characteristics of MnTe-x%Cr samples. M-H characteristics are measured within the range of -12T to 12T at different temperatures in the FM/AFM and PM regimes. As expected, all MnTe-x%Cr samples, except MnTe, show saturation at higher fields. For MnTe, no saturation is observed. MnTe shows a different M-H trend without any hysteresis loop in the -12T to 12T range within the AFM temperature range. In general, magnetic moment (emu/mole) increases with the increase of Cr doping due to the presence of strong FM Cr-Cr bonding and weak AFM Mn-Cr bonding. Both Cr-Cr and Mn-Cr bonds reduce the moment of the AFM Mn-Mn bond. All the FM samples show hysteresis characteristics with certain coercivity and retentivity (listed in Table 1). The residual magnetic moment of MnTe-x%Cr samples ($x>0$) increases with $x$. All samples except MnTe-14%Cr and MnTe-20%Cr show linear M-H relation in the paramagnetic temperature regime. Both MnTe-14%Cr and MnTe-20%Cr show FM behavior in the paramagnetic temperature range, which can be attributed to the existence of FM polarons due to the short-range ferromagnetic correlation that can exist above the transition temperature [30].

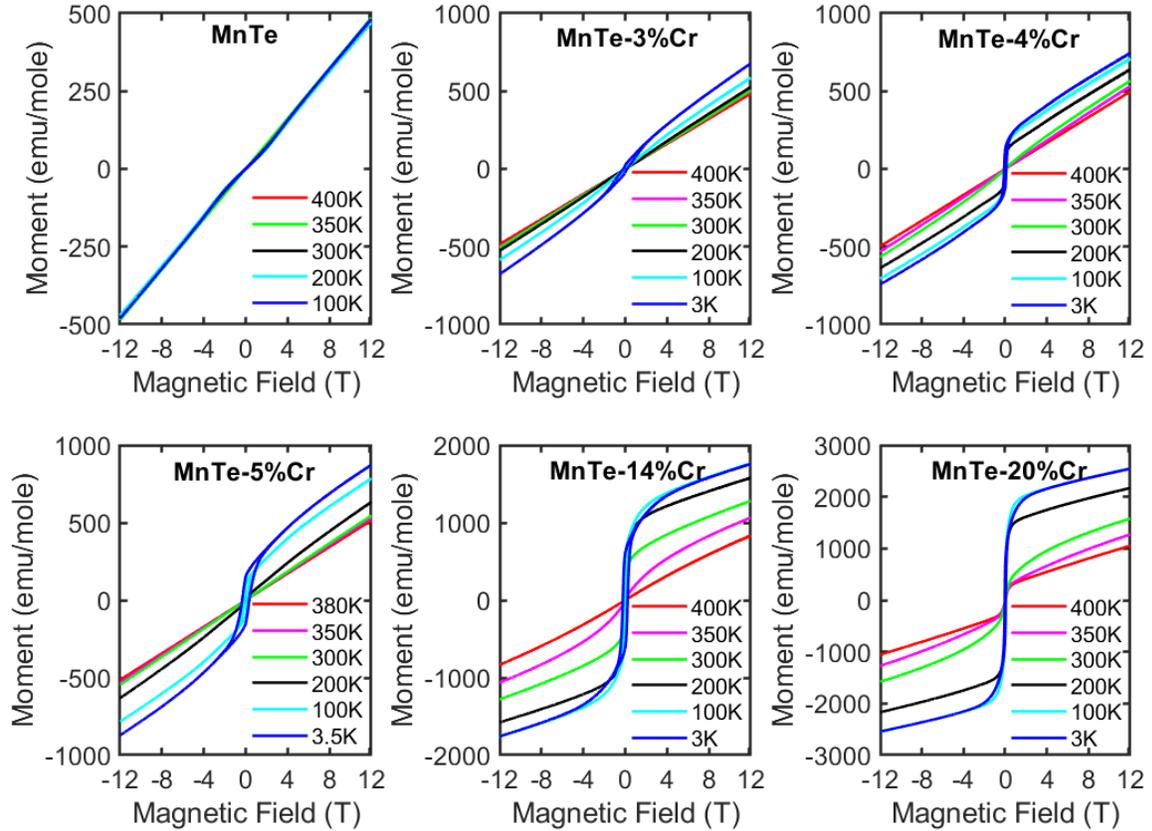

Figure 4: Magnetic moment versus magnetic field at different temperatures for MnTe-x%Cr samples.

Both magnetic moments versus temperature and magnetic field demonstrated that Cr-doping introduces ferromagnetism in antiferromagnetic MnTe. Ferromagnetism in MnTe-x%Cr can have three possible explanations; magnetic frustration due to the competition between FM and AFM exchange interaction, deviation of the magnetic structure from AFM to a canted-spin structure, and the spin-polarized hole-mediated ferromagnetic interaction [19,23,24,31]. When Cr ions with the high spin configuration of $d^4$ substitute Mn ions, $p$-$d$ hybridization between $p$-orbitals of Te atom and the $d$-orbitals of Cr introduces spin-polarized holes, which can control the FM coupling among neighboring Cr ions [24]. In MnTe, AFM behavior comes from the spin-configuration of Mn ions where Mn $d$-orbitals of majority-spin (spin-up) are fully occupied with the high-spin configuration of $d^5$ ($Mn^{2+}$) and the minority-spin (spin-down) $d$-orbitals are empty [19,24].

The spin-spin interaction is expected to be dependent on the location of the Mn $d$-orbital with respect to the valence band maxima [24]. The $p$-$d$ and $s$-$d$ exchange interaction between carriers and spins plays a vital role in determining the magnetic properties of dilute magnetic semiconductors (DMS) [24]. The net value of the Mn-Mn exchange constant within the nearest-neighbor shell is mainly determined by a hole-induced ferromagnetic interaction and the intrinsic short-range antiferromagnetic superexchange interactions [19,24]. At the low amount of Cr concentration, the introduction of ferromagnetic interaction dominates over the antiferromagnetic ground state. With the increase of Cr, the antiferromagnetic ordering in the system is completely lost [19]. Cr doping can also deteriorate the *sp-d* exchange, which can reduce the metallic behavior of the system [19].

Besides the hole-controlled FM interaction, canted FM structure into AFM may also exist in the bulk samples where substitution of Mn-ions by Cr-ions with $d^3$ configuration can create small clusters of FM phases. Therefore, the inclusion of Cr can provide manifold modifications in the host Mn-Mn AFM bonding (Figure 3), i.e., Cr can create weak Mn-Cr AFM/FM bonding, Cr can introduce Cr-Cr FM bonding, which should be less probable in lower Cr concentration, or Cr can modify the bond length and angle, which can create a canted FM structure into AFM ordering. Based on the magnetic properties, it can be concluded that both FM and AFM phases exist in Cr-doped MnTe samples, where FM clusters have a dominating magnetic moment over the AFM moment of MnTe.

**Heat Capacity Trends in Cr-doped MnTe: Evidence for Magnons**

Thermodynamically, thermopower is the amount of the entropy carried by the charge carriers. Heat capacity is considered as one of the effective methods to trace the existence of different entropy carriers, namely phonons from the quantized wave of lattice vibration, dilation from thermal expansion of lattice, Schottky from the lifting of the degeneracy of ions, and magnons from the quantized spin-wave [32]. Heat capacity is also a sensitive measurement of magnetic phase impurities, which can cause anomalous features in heat capacity due to the formation of localized moments [33]. Therefore, heat capacity measurements were performed on MnTe-5%Cr and MnTe-14%Cr along with undoped MnTe to probe the existence of magnons and other possible magnetic phase impurities. Figure 5 illustrates the comparison of the specific heat capacity (Cp) for the undoped, 5%, and 14% Cr doped MnTe.

In MnTe, heat capacity contributions are primarily due to the lattice, dilation, Schottky, and magnons [32]. Magnon heat capacity of MnTe shows a sharp peak at around ~307K, which is also the Néel temperature of MnTe. Interestingly, all the Cr-doped MnTe samples also show a similar peak at around the same temperature, with a slight reduction in the peak value. To determine the magnon heat capacity contribution in Cr-doped MnTe, we have theoretically determined the non-magnonic heat capacity contributions from lattice ($C_v$), dilation ($C_d$), and Schottky ($C_{Sc}$) based on the parameters of MnTe (shown in Figure 5) [32]. From the heat capacity trends of Cr-doped MnTe, non-magnonic heat capacity contributions are expected to be the same for all the samples because of low doping concentration, the insignificant contribution from Schottky (<3% of total heat capacity [32]) and iso-structure of MnTe and CrTe. Therefore, we can state that the difference in the heat capacity between doped and undoped MnTe is coming from the difference in magnon heat capacity contribution. The reduction in magnon heat capacity of Cr-doped MnTe can be attributed to the suppression of the antiferromagnetic magnons caused by the weak Mn-Mn exchange interaction due to the low-spin (LS) state of $Mn^{2+}$ in Cr-doped MnTe. In undoped MnTe, $Mn^{2+}$ ions have a high-spin (HS) state (shown in Figure 5). The spin state and spin number of $Mn^{2+}$ in undoped and doped MnTe are also confirmed by the magnetic measurements. The decline of the magnon contribution into heat capacity can also be caused by the decrease in the AFM MnTe phase due to the introduction of the FM phase by Cr-doping. As no other peaks appear at FM transition temperatures of Cr-doped MnTe, this suggests that the broad and large peak of AFM magnon is masking the small FM magnon contribution.

The Field-dependent heat capacity of the Cr doped MnTe is performed at 12T fields and demonstrated in Figure 5. Field-dependent heat capacity can sometimes illustrate the presence of both AFM and FM magnons. AFM and FM magnons show different trends depending on the thermal energy and material properties like spin-wave stiffness, exchange energy, spin number, and magnetization [1]. Under the magnetic field, heat capacity from AFM magnons mostly remains unchanged due to its high spin-flop field. In contrast, FM magnons can be quenched by a high magnetic field depending on the material properties and the temperature [1,34]. As those particular properties were not available for the materials under study, we performed the field-dependent heat capacity measurement in the temperature range of interest to check for any differences or the quenchable FM magnons. Here, a more important note is that there are some possible Cr-based impurity phases apart from CrTe ($T_C$~342K), such as $Cr_2Te_3$ ($T_C$~180K), $Cr_5Te_8$ ($T_C$~220K), and $Cr_3Te_4$ ($T_C$~330K). Though we did not see any traces for them in the XRD data, they could still appear in the heat capacity if they existed in small quantities. This is especially important as the magnon-bipolar carrier drag exists at a range of temperatures that overlaps with possible contributions from the low- or high- temperature FM impurity phases if they lived in the system. $Cr_2Te_3$ and $Cr_5Te_8$ have $T_C$ far from $T_N$ of MnTe, and the heat capacity data does not indicate the presence of FM magnons from these phases. However, CrTe and $Cr_3Te_4$ have $T_C$ near $T_N$ of MnTe. The field-dependent heat capacity data, however, does not indicate any noticeable impacts on the AFM magnons if such phases existed in the system. Therefore, we can ignore contributions from such potential FM phases to the magnon-bipolar carrier drag studied here.

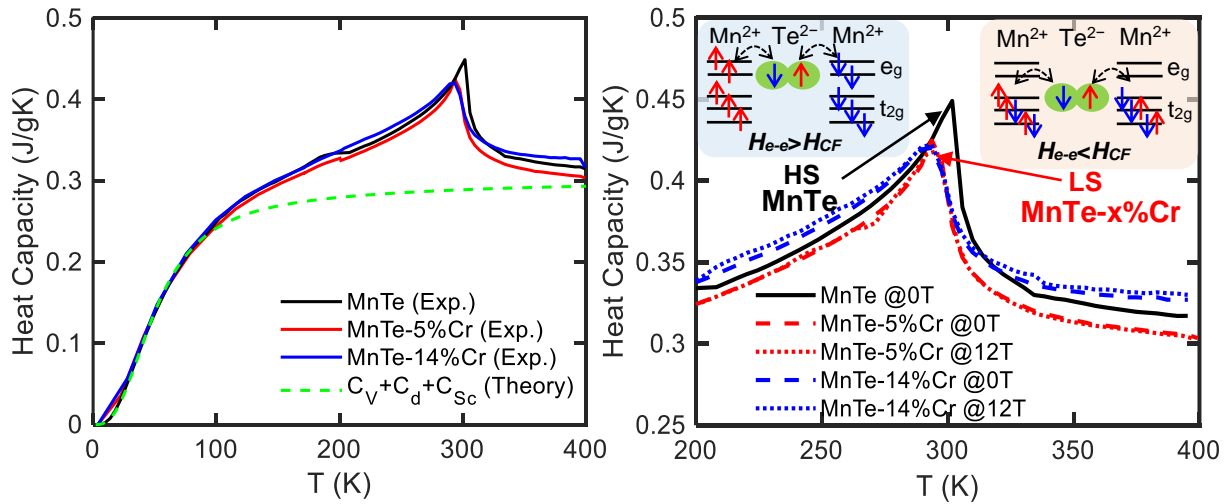

Figure 5: (Left) Heat capacity ($C_p$) of MnTe, MnTe-5%Cr, and MnTe-14%Cr from experiment along with estimated lattice, dilation and Schottky heat capacity contributions for MnTe. (Right) $C_p$ of MnTe at 0T, MnTe-5%Cr and MnTe-14%Cr at 0T and 12 T magnetic field within 200K-400K range with the illustration for HS and LS electronic configurations in MnTe and MnTe-5%Cr, respectively.

Both magnetic susceptibility and heat capacity trends of Cr-doped samples suggest the presence of both FM and AFM clusters with insignificant contributions from other possible magnetic phase impurities. As in Cr-doped MnTe, long-range ordering is still dominated by the AFM ordering;

similar magnonic contribution into Cp with reduced peak value was observed in MnTe-x%Cr samples. The heat capacity analysis for Cr-doped MnTe samples implies that the major difference in the heat capacity of MnTe-x%Cr compared to MnTe is due to the differences in the AFM magnon system. Both the lower exchange interaction energy in MnTe-x%Cr due to LS $Mn^{2+}$ and FM-AFM clustering can cause a reduction in the magnonic heat capacity. The quantitative analysis of MnTe-x%Cr requires separate attention, which is beyond the scope of this article. At higher temperatures (>400K), Cp is approximately constant at around 0.28 J/g-K according to the Dulong-Petit limit.

**Hall Properties of Cr-doped MnTe: Evidence for Bipolar Nature**

To determine the carrier transport nature in Cr-doped MnTe, we performed the Hall measurements using Van-der-Pauw method on pristine MnTe, 4% Cr-doped, and 20% Cr-doped MnTe samples under 0 T and 12 T magnetic field. Sheet resistances were calculated from the data at 0 T, and the Hall coefficients ($R_H$) were obtained under 12T. Hall carrier concentration is directly derived from RH using the relation of $n_H = 1/{eR_H}$, where $e$ is the charge of carriers. This relation gives the net carrier concentration and is applicable to both single and two carrier systems; hence, it can demonstrate both the intrinsic and extrinsic carrier nature of the materials. The type of Hall carrier is determined from the sign of the Hall voltage. The carrier concentration obtained from the Hall measurement is illustrated in Figure 6. Hall mobility can easily be calculated from the resistance and carrier concentration for a single carrier type system, but the calculation of Hall mobility for the intrinsic system requires carrier effective mass information. Therefore, the estimation of the Hall mobility for 4%Cr-doped MnTe using a single carrier model can be misleading. Since the carrier mobility is not an essential factor to support the theoretically formulated magnon-bipolar carrier drag thermopower, we have discussed it in the appendix.

As shown in Figure 6, MnTe shows p-type dominant carrier conduction, and 20%Cr-doped MnTe shows n-type dominant carrier concentration. Hall parameters of the pristine MnTe follow the trends reported in [13]. 4%Cr-doped MnTe has lower carrier concentration with a temperature-dependent increment trend, which is attributed to the intrinsic carrier nature. The bipolar carrier concentration and electrical conductivity generally increase in intrinsic materials due to the thermal excitation of electron-hole pairs. However, in the case of doped samples, the dopants with carriers of opposite charge can compensate for the carrier contribution from the host leading to less carrier concentration and intrinsic nature.

Similarly, Cr-doping is not enhancing the number of carriers in low Cr-doped MnTe but makes the sample less p-type by compensating the acceptor centers and reducing the number of holes. In other words, Cr is neutralizing some p-type acceptors to provide overall a smaller number of carriers. As such, the samples with low Cr doping approach the intrinsic regime, where the bipolar carriers (electron-hole pairs) due to the thermal excitation are the dominant carriers. At high enough Cr doping, the sample becomes n-type with electrons dominating the transport properties. Overall, the p-type electrical conductivity of MnTe decreases with the increase of Cr-doping while n-type conductivity is increased.

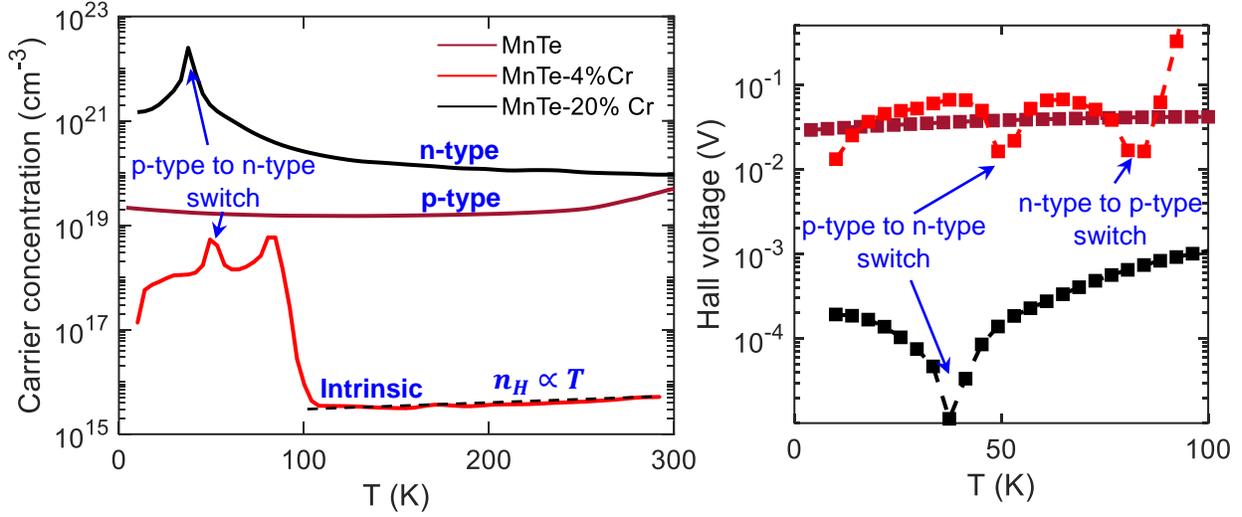

Figure 6: (Left panel) Temperature-dependent carrier concentration of MnTe-x%Cr samples (x=0, 4, and 20), (Right Panel) the absolute value of the Hall voltage showing the switching of carrier type between p and n-type.

Both carrier concentration and Hall voltage of MnTe-20%Cr show a discontinuity at around 40K, which indicates the p-type to n-type conversion. Similar types of carrier concentration switching were reported in previous literature [35]. After the conversion, MnTe-20%Cr remains n-type with a higher carrier concentration than MnTe. Interestingly, MnTe-4%Cr shows two discontinuity points in both carrier concentration and Hall voltage at around 50K and 85K. The former can be attributed to p-type to n-type conversion, while the latter indicates switching back from n-type to p-type above 85 K. This type of Hall voltage trend is also an indicator of the electronic transport contributions from both n-type and p-type carriers; hence, the near intrinsic nature of the sample [36]. Once the electrons donated by Cr ions compensate the holes from the acceptor centers in MnTe, the overall carrier concentration drops rapidly. It then increases linearly (in log scale) with the temperature, which is attributed to the exponential thermal excitation of the bipolar carriers. The sharp drop of the carrier concentration in MnTe-4%Cr evidences the hypothesis that Cr-doping compensates the acceptor centers of MnTe by introducing n-type carriers; hence, resulting in an intrinsic nature into low Cr-doped MnTe. It should be noted that the intrinsic Mn-vacancies in MnTe and the multiband structure of Cr-doped MnTe can lead to complex carrier transport as well as different scattering mechanisms along with the anomalous Hall effect due to the presence of ferromagnetism which is beyond the scope of this study.

**Thermoelectric Properties of Cr-doped MnTe: Evidence for Magnon-bipolar Carrier Drag**

According to the earlier discussion of the magnetic susceptibility, heat capacity, and Hall properties, a low Cr-doped MnTe is a good material system for studying the bipolar carrier drag by magnons.

To investigate the impact of the magnon-bipolar carrier drag on the thermoelectric transport properties of Cr-doped MnTe samples, we measured their temperature-dependent electrical resistivity, thermal conductivity, and thermopower. The electrical conductivity and thermopower of MnTe-x%Cr are shown as a function of temperature in Figure 7, which shows a good agreement with previously published data [18] as well as Hall properties shown in Figure 6.

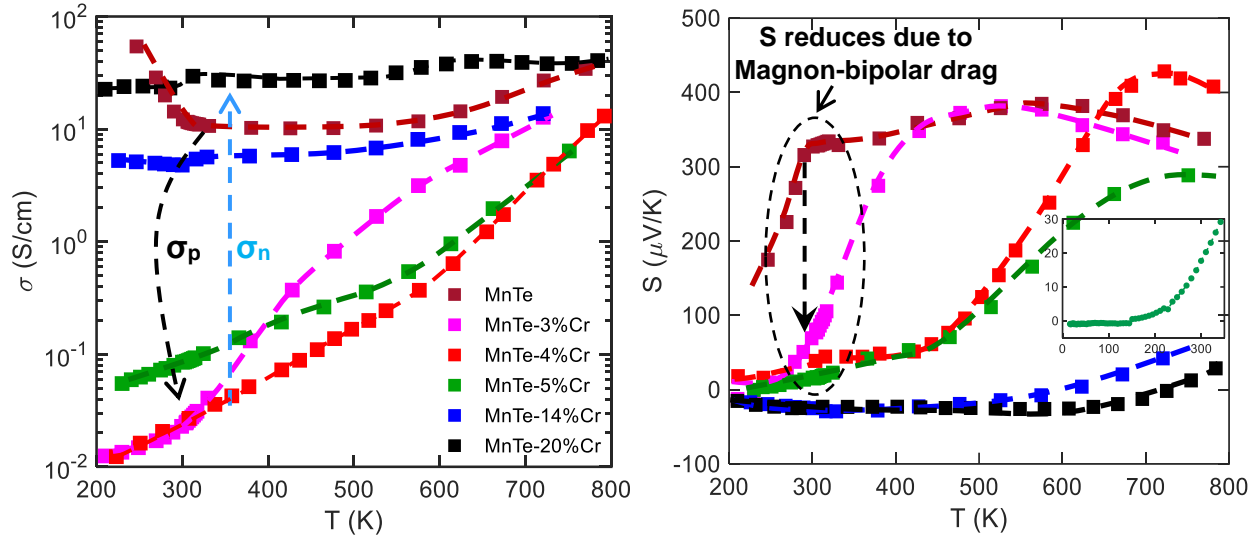

Figure 7: Electrical conductivity (left) and thermopower (right) of different MnTe-x%Cr samples as a function of temperature. Inset shows low temperature thermopower (10-350K) of MnTe-5%Cr. Both conductivity and thermopower demonstrate the impact of bipolar effects.

According to the electrical conductivity trends, Cr-doping reduces the p-type conductivity ($\sigma_p$) of MnTe due to the $Cr^{3+}$ oxidation state, which makes Cr a donor type dopant [19,25]. As mentioned earlier, Cr-doping reduces total carrier concentration by compensating the acceptor centers and hence reduces the electrical conductivity. After the initial reduction of electrical conductivity in the MnTe-3%Cr sample, electrical conductivity increases with the increase of Cr-doping due to the rise in n-type conductivity ($\sigma_n$) and conductivity of MnTe-20%Cr becomes higher than that of MnTe. This trend is also supported by the Hall data, where it is observed that n-type carrier concentration increases with the increase of Cr-doping and MnTe-20%Cr has a higher carrier concentration than that of MnTe. Cr-doped MnTe samples exhibit semiconducting nature in electrical conductivity, while MnTe shows an extrinsic to intrinsic transition in electrical conductivity at around Néel temperature. Saturation of the conductivity in MnTe after transition temperature can be attributed to the spin-disorder scattering of the carriers [37]. n-Type carrier nature of Cr-doped MnTe samples is observed in the thermopower data. In general, the electrical conductivity of Cr-doped MnTe is suffered from bipolar effect.

Temperature-dependent thermopowers of MnTe-x%Cr samples demonstrate a significant reduction due to the bipolar effect at lower temperatures. In contrast, that of the MnTe sample shows a small drop due to bipolar effect only at higher temperatures. Thermopower of undoped MnTe shows a distinct magnon-drag contribution below the Néel temperature (~307K); it increases with temperature, saturates at $T_N$, and continues growing with a small slope. The extension of the excess thermopower above the $T_N$ has been attributed to the paramagnon-drag effect [13]. The thermopower starts declining at around 550K due to the temperature-induced bipolar effect. The thermopower trend in Cr-doped MnTe is different than that of MnTe due to the magnon-bipolar carrier drag effect. They show almost no sign of magnon-drag below or at around

300 K, where the magnon heat capacity shows the maximum value. Both carrier concentration and electrical conductivity data confirm the bipolar semiconducting nature in Cr doped MnTe samples. The heat capacity confirms the presence of magnon at around 300K. Therefore, the reduction in total thermopower observed in Cr doped MnTe samples supports the concept of magnon-bipolar carrier drag effect. The decline is also in agreement with the theoretical formulation.

The magnon-bipolar carrier drag effect is similar to the phonon-bipolar drag effect (illustrated in Figure 1). Here, the sign of the magnon-drag thermopower is dependent on the type of the carriers dragged by the magnon. Magnon-hole drag provides a positive contribution to the total thermopower, while magnon-electron drag provides a negative contribution. Therefore, the overall thermopower performance of Cr doped MnTe deteriorates below and around the transition temperature. MnTe-3%Cr shows a small positive thermopower at around 300K with almost no amount of net magnon drag thermopower. MnTe-4%Cr and MnTe-5%Cr show nearly zero thermopowers at approximately 200K, which extends up to 10K (see inset of Figure 7(right)). The magnon-bipolar effects have suppressed both the diffusion and drag thermopower in these samples. In the low-temperature region, Hall voltage also changes sign a couple of times for the low Cr-doped MnTe sample, which is also the indicator of the near intrinsic transport and the bipolar carrier conduction. The Hall voltage is sensitive to the carrier type. It goes back and forth between the hole- and electron-dominant voltage presumably due to the changes of the Fermi energy or the band structure when the Fermi is near the intrinsic level. The thermopower is too small, and the measurement sensitivity is not adequate to display such variations. Overall, MnTe with lower Cr-doping shows reduced thermopower due to a more balanced electron and hole magnon-bipolar carrier drag where the Hall data indicate two-carrier conduction.

Interestingly, MnTe-14%Cr and MnTe-20%Cr show small negative magnon-electron drag thermopower at around 300 K (shown in Figure 8), which indicates the increase of magnon-electron drag over magnon-hole drag. The higher Cr-doped MnTe samples show negative thermopower at lower temperatures due to the dominant n-type transport. These trends demonstrate the existence of the magnon-bipolar drag effect. Here, it is essential to note that magnon-drag in MnTe-x%Cr is caused by the AFM magnon from Mn-Mn interaction, which maximizes at around 300 K. Heat capacity data shows insignificant or near-zero FM magnons contribution from FM exchange interactions. Therefore, there is a loose relationship between the observed magnon-carrier drag thermopower in MnTe-x%Cr and their corresponding magnetic transition temperatures, which are associated with their FM nature discussed in the earlier section. Hence, the temperatures associated with the peak of the magnon-carrier drag thermopower in MnTe-x%Cr can be different from their respective Curie temperatures. For MnTe-14%Cr and MnTe-20%Cr samples, negative thermopower continues until up to around 650 K and 730 K, respectively. The sign change for the thermopower of $Mn_{1-x}Cr_xTe$ (x=0.14 and 0.20) samples occurs due to the excitation of the carriers at higher temperatures, which introduces temperature-induced bipolar effect on thermopower, as Cr-doping reduces the bandgap of MnTe (refer to the bandgap measurement analysis in appendix). Interestingly, for the low-doped MnTe samples (%Cr<6), thermopower at paramagnetic domain follows the trends of MnTe, and at high temperatures, thermopower reaches the value of thermopower of the pristine MnTe. The shift of the thermopower rise to higher temperatures among different doped samples correlates with their hole concentrations (and how close they are to the intrinsic regime), i.e., the lower the hole concentration (or, the higher the intrinsic character), the rise is shifted to a higher temperature, where the contribution from the thermally excited holes takes over the trend of the thermopower.

In general, with the increase of Cr-doping, thermopower becomes more negative at lower temperatures, and its magnitude is reduced at higher temperatures.

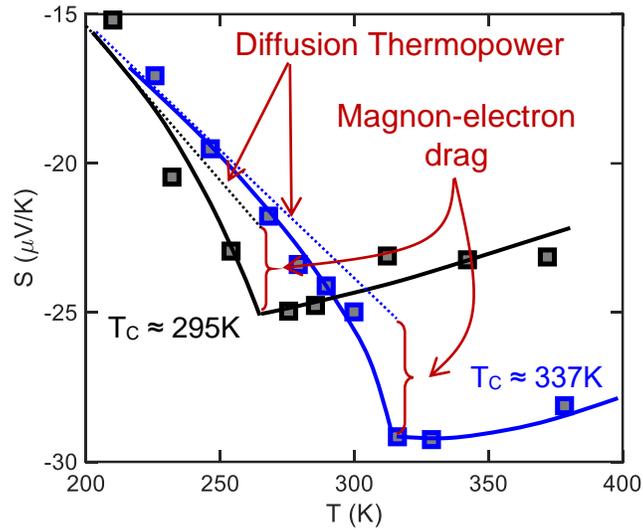

Figure 8: Magnon-electron drag thermopower of MnTe-14%Cr (blue) and MnTe-20%Cr (black) samples.

Like electrical conductivity and thermopower, bipolar conduction can also affect thermal conductivity (shown in Figure 1). Figure 9 presents the thermal conductivity of Cr-doped MnTe samples. The thermal conductivity of MnTe initially reduces upon 3% Cr doping and then increases with the increase of Cr doping, which can be attributed to the bipolar conduction effect. A more detailed theoretical study is needed on thermal conductivity, as it has contributions from phonons, charge carriers, and magnons. Theoretical study to determine the contribution of each component is, however, beyond the scope of the current study. According to the literature [38], near the transition temperature, the scattering of phonons by spin disorder is strong; however, the magnon contribution to the total thermal conductivity is small due to the small magnon mean free path of MnTe. According to electrical conductivity and carrier concentration, electronic contribution to the total thermal conductivity increases with the increase of the Cr-doping as well as the temperature, which can at least partially explain the rise of the total thermal conductivity versus Cr-doping by attributing to the bipolar carrier conduction effect. The initial reduction in thermal conductivity with low Cr doping (3% at) may be attributed to the introduction of impurity scattering and/or alloy scattering, which may not be increased significantly with the increase in Cr-doping. The thermal properties of the MnTe samples are mainly dominated by the lattice component through phonon dynamics [39]. Overall, thermal diffusivity, as well as thermal conductivity, reduces with increasing the temperature due to a higher three phonon scattering at elevated temperatures.

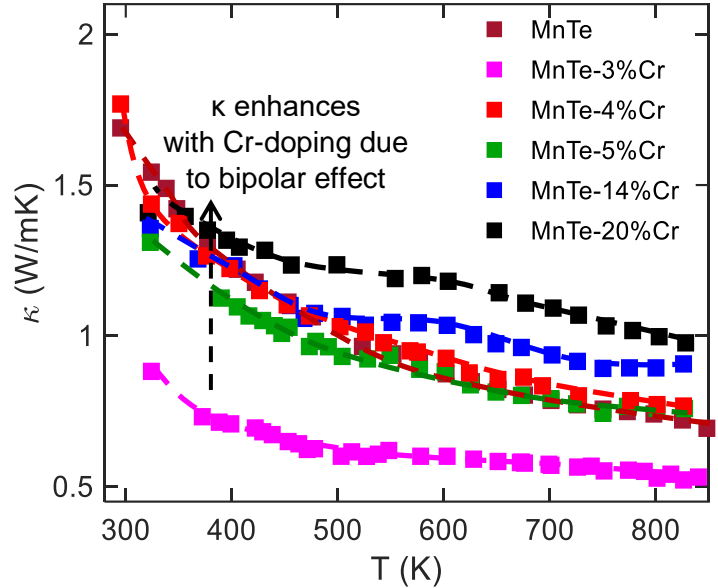

Figure 9: Thermal conductivity of MnTe-x%Cr samples as a function of temperature.

Finally, the thermoelectric power-factor and figure of merit (zT) are calculated from the measured quantities, which also suffer from the magnon-bipolar carrier drag effect. Thermoelectric power factor (PFT) is the product of the square of thermopower ($S$), electrical conductivity ($\sigma$), and temperature ($T$) ($PFT = S^2\sigma T$). On the other hand, the thermoelectric figure of merit is a dimensionless quantity that measures the performance of the thermoelectric materials from the material properties. Thermoelectric figure of merit is defined by $zT = \frac{S^2\sigma}{\kappa}T$. Figure 10 compares the power factor and figure of merit of the samples, respectively. As expected, they both follow similar trends as those of the electronic and thermal properties of the samples. MnTe has a higher power factor as well as higher zT among all samples. Among other samples, 3% Cr doped MnTe shows better thermoelectric behavior. With the increase of Cr content, thermoelectric behavior degrades significantly due to loss of carrier concentration and the bipolar magnon carrier drag. Both MnTe-14%Cr and MnTe-20%Cr show an abrupt change in the power factor and zT due to the thermopower sign change from n-type conductivity to p-type. The reduction in the thermal conductivity of $Mn_{0.97}Cr_{0.03}Te$ did not improve its zT due to the dominance of the bipolar effects in both electrical conductivity and thermopower, which is consistent with the previous reports [40].

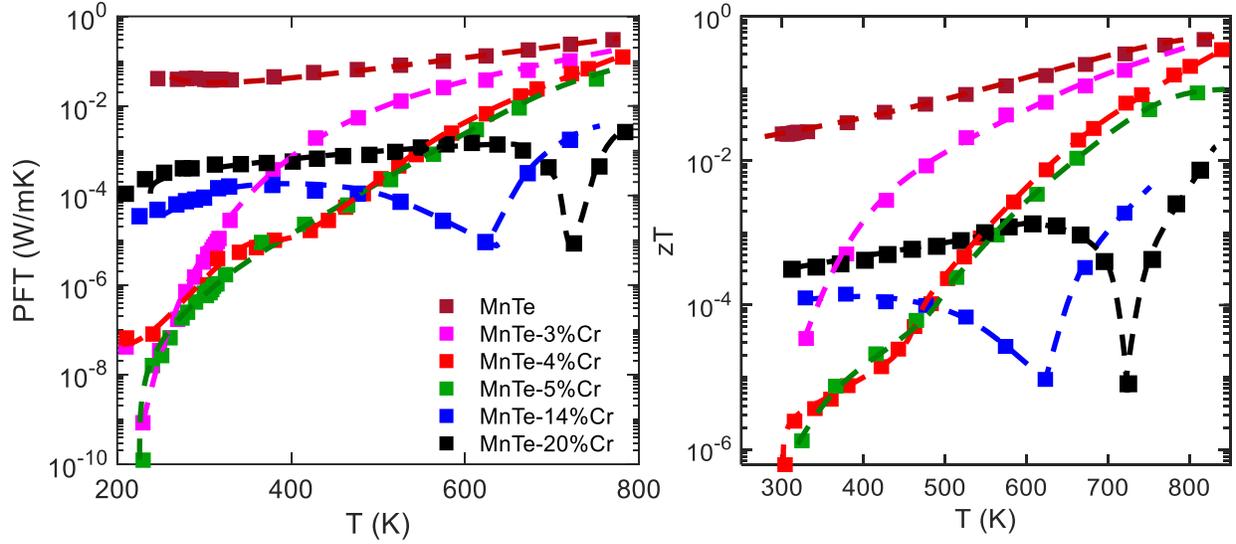

Figure 10: Thermoelectric power factor times temperature (PFT) and thermoelectric figure-of-merit (zT) as a function of temperature for different MnTe-x%Cr samples.

**Experimental evidence for Relaxation Time Independent Magnon-bipolar Carrier Drag Thermopower in Intrinsic Magnetic Semiconductors**

Besides the prediction of the reduction of total thermopower due to the magnon-bipolar carrier drag, the theoretical model also predicts that the total thermopower for the intrinsic magnetic semiconductor with magnon-bipolar carrier drag is independent of the carrier and magnon relaxation times. To verify this prediction experimentally, we made two MnTe-5%Cr samples with different densities of 5.97g/cm$^3$ and 5.82 g/cm$^3$, which are 99.5% and 96.9% of the ideal density, respectively. As discussed in the sample preparation section, a low-density sample was sintered at 950°C with a short soaking time of 10 mins, and under a lower pressure of 17MPa. The high-density sample was soaked for 20 min under higher pressure of 50MPa. The low-density sample can experience smaller grain growth during sintering and less healing of defects at the grain boundaries. Therefore, the low-density sample has a higher density of interface defects, which would impact the carrier and magnon relaxation times. Thus, the associated carrier and magnon relaxation times are expected to be significantly different in the two samples due to the large difference in their crystal defects. The variations of the crystal defects may also lead to a variation of the scattering exponent; however, this change must be insignificant due to the similarity of the scattering mechanisms. Therefore, per equ. (11), the two $Mn_{0.95}Cr_{0.05}Te$ samples must provide similar thermopower trends, due to the insignificant changes in the carrier concentration and the scattering exponent, despite the significant changes in the relaxation times. Note that at low temperatures, the carrier generation from thermal excitation is insignificant; therefore, the carriers' concentrations, as well as their effective masses, remains unchanged, which satisfies the conditions in (11). The electrical conductivity and thermopower were measured as a function of temperature for the two MnTe- 5%Cr samples as compared in Figure 11 along with other low Cr-doped samples (MnTe-3%Cr and MnTe-4%Cr).

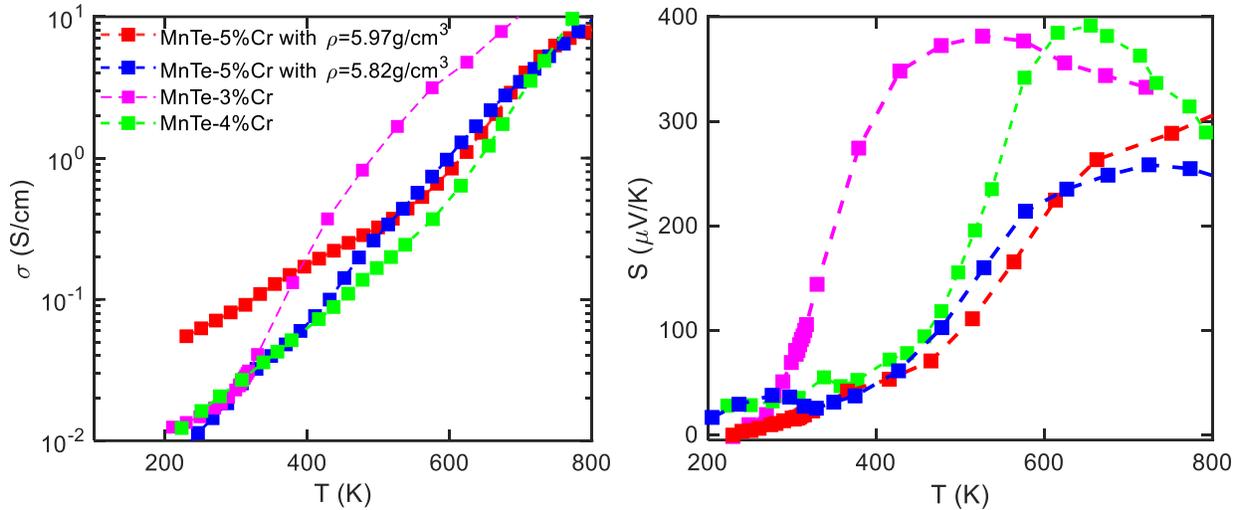

Figure 11: Comparison of the electrical conductivity and thermopower of low and high-density MnTe-5%Cr samples along with 3%Cr and 4%Cr doped MnTe.

The temperature and soak time can affect the dopant substitution in general. The data show different trends for the electrical conductivity of 3%, 4%, and 5% doped samples. Among them, the 3% doped sample has the highest p-type conduction evidenced by its higher electrical conductivity and higher thermopower data. The dopant compensation is higher in the 4% doped sample, i.e., it has less hole concentration than the 3% doped sample. This sample has still p-type conduction but at a lower concentration, as evidenced by its thermopower, which remains low due to bipolar conduction up to a higher temperature. Above approximately 500 °C, the contribution from the excited holes dominates over that of the electrons, and the thermopower increases. The 5% doped sample shows even less p-type conduction as the thermopower remains low up to higher temperatures compared to those of the 3% and 4% doped samples. The electrical conductivity of the 4%Cr and low-dense 5%Cr dense samples are comparable within the experimental tolerances and the variations of the carrier mobility of different samples. Due to the similar conductivity above 500K, it can be concluded that the low density 5%-doped MnTe sample has a similar carrier concentration as the high-density sample while having a higher defect scattering near and below the room temperature where the magnon-bipolar carrier drag exists and shows a carrier relaxation time-independent nature.

At high temperatures, where phonon scattering is dominant, both samples show almost similar conductivity. At 300K, where the magnon-drag exists, the difference in the conductivities is more than three times. Assuming same carrier concentration for both samples, the difference in the electrical conductivities is coming from the difference in the carrier relaxation times. It can be seen that while the carrier relaxation times are ~3X different in the two samples, the thermopowers are close within the experimental tolerances, which agrees well with the predicted low sensitivity of the bipolar magnon drag to the relaxation times. Note that both samples have the same elemental composition; therefore, the difference in the relaxation times must be arising due to the scattering from defects and grain boundaries. As expected, the thermopowers of the two samples are closely matched and follow the same trend below 600K. This trend evidence two findings: (i) insignificant impact of the grain boundary scattering on magnon drag thermopower, and (ii) the negligible effect of the relaxation time on the total bipolar thermopower supporting equ. (11). Indeed, the first conclusion has also been reported for the case of the FM metals by Blatt et al. in 1967 [41] and

Watzman et al. in 2016 [2]. They each showed that grain boundary scattering in FM metals (Fe first, then Ni and Co) affects thermopower only by suppressing phonon drag, with no impact on the diffusion or magnon drag contributions.

**Conclusion**

We presented the evidence for the existence of the magnon-bipolar carrier drag effect in intrinsic magnetic semiconductors. Magnon electron drag offers effective means to enhance the classical diffusive thermopower in the magnetic materials for thermoelectric applications. Magnon carrier drag in metals and degenerate semiconductors can add a significant advective thermopower to the diffusive one near the transition temperature – a process that has been commonly reported in the literature. However, magnon carrier drag in intrinsic semiconductors has not been studied as much. We showed both theoretically and experimentally that in such a case, the bipolar transport can adversely affect the overall magnon carrier drag thermopower. The thermopower expression for a bipolar magnetic semiconductor was formulated based on the diffusion and drag thermopower and the electrical conductivity of the electron and hole and found to be independent of both the carrier and the magnon relaxation times. For the proof of concept, detailed experiments were carried out on MnTe-x%Cr compounds with $x = 0, 3, 4, 5, 14$, and $20$. Thermoelectric transport properties, Hall properties, heat capacity, and magnetic susceptibility measurements were performed to investigate the theoretically predicted conclusions for magnon-bipolar carrier drag thermopower. The chemical, optical, magnetic, electrical, and thermal transport properties of the Cr doped MnTe samples are studied from sample characterizations, including XRD and UV-Vis. The magnon-bipolar drag was observed for all samples with a distinct magnon-electron drag component for higher Cr-doped samples. Due to the reduction of the hole concentration and the thermopower, Cr doping lowered the thermoelectric figure of merit despite the decrease of the thermal conductivity. Due to the co-existence of both ferromagnetic and antiferromagnetic phases, the heat capacity showed a magnon contribution like that of MnTe, while magnetic susceptibility demonstrated the existence of the dominating ferromagnetic phase only. The clustering of FM and AFM phases in MnTe-x%Cr samples can further modify the magnon-carrier interactions.

**Acknowledgments:**

Authors acknowledge the funding support by the Air Force Office of Scientific Research (AFOSR) under contract number FA9550-12-1-0225 and the National Science Foundation (NSF) under grant numbers ECCS-1351533, ECCS-1515005, and ECCS-1711253.

**Appendix**

**Sample Preparation**

MnTe-x%Cr samples with $x = 0, 3, 4, 5, 14$, and $20$ were synthesized using 99.99% pure Mn, Te, and Cr powders along with a pristine MnTe control sample. Elemental powders were milled under the Ar environment in a tungsten carbide cup for 8 hrs at 650 rpm using a Fritsch P7PL planetary ball mill. The powders were subsequently annealed at 850°C for 24 hrs under vacuum in a rocking furnace. The annealed samples were crushed and milled again for 8 hrs. A homogenous MnTe phase without the general phase impurity of MnTe$_2$ was achieved. The powders were loaded into graphite dies with a 6mm hole diameter inside an atmosphere-controlled glove box filled with Ar, and subsequently consolidated into cylindrical ingots using a spark plasma sintering (SPS) equipment located inside the same glove box. The SPS operation inside the glovebox ensures oxygen absence and prevents the highly probable reaction $2MnTe + ½ O_2 \rightarrow MnTe_2+MnO$. Spark plasma sintering was performed at approximately 50 MPa pressure at a constant heating rate of 60 °C/min at a maximum temperature of 950 °C and the soaking time of 20 mins. $O_2$ and $H_2O$ levels were always kept at < 0.01 ppm inside the glove box. All consolidated ingots had >97% density of the ideal value. To demonstrate the relaxation time-independent thermopower in the intrinsic magnetic semiconductor, we have synthesized low-density MnTe-5%Cr ($\rho$ = 5.82 g/cm$^3$) with short soaking time (10mins), under lower pressure (17MPa), at 950°C. The short soak time leads to smaller grain growth; consequently, it may result in smaller grains size to enhance the crystal defects intentionally.

**Characterization of the Samples**

*X-ray diffraction (XRD)* patterns were collected using Rigaku Miniflex with Cu-Kα radiation at 0.154 nm wavelength and analyzed with PDXL software version 2.8.4.0 to extract different lattice parameters. All the XRD patterns are measured from a disk sample of 6mm diameter at a rate of 5°/min within the range of 10°-90°.

The *optical bandgap* of MnTe-x%Cr is measured with the optical spectroscopy using a UV-Vis spectrometer (Evolution 200) with a Xe light source and wavelength range of 190 nm – 1100 nm. The absorption coefficient (α) for the direct transition was measured for the powdered samples using an integrating sphere.

*Magnetic susceptibility* was measured versus temperature and magnetic field with the vibrating sample magnetometry (VSM) method using the Quantum Design DynaCool 12T system. Temperature-dependent magnetic susceptibility (M-T) was measured from 2K to 400K at 1000 Oe under a vacuum environment. The susceptibility versus magnetic field (M-H) was measured at selected temperatures over -120 kOe to 120 kOe.

The *specific heat capacity* ($C_p$) was measured using the Quantum Design DynaCool 12T heat capacity option from 2K-400K. Heat capacity for higher temperature was assumed based on the Dulong-Petit (DP) limit.

*Hall measurement* was performed with the Van-der Pauw (VdP) method using Quantum Design DynaCool 12T. A thin disk (diameter 6 mm, thickness <0.6 mm) is cut from the ingot and measured the Hall data under 0T to obtain longitudinal resistivity and under 12T to obtain Hall resistivity. Longitudinal resistivity from VdP data is extracted numerically using zero-finding technique. Carrier concentration is calculated from the inverse of Hall coefficient times charge. Hall mobility is estimated using the single carrier formula.

*Electrical conductivity* and *thermopower* were measured simultaneously with the standard 4-point probe method using Linseis equipment under the He environment from 200K to 850K. The original Linseis software calculates the thermopower from a single temperature difference (ΔT) and voltage difference (ΔV), which is quite often erroneous. Therefore, the measurement was performed for five different temperature differences, and each measurement was repeated four times and then averaged. The thermopower was calculated from the slope fitting to five separate temperature and voltage differences. The accuracy of the measurement was verified by inspecting the linear fit to the (ΔV-ΔT) data set.

The *thermal diffusivity* (υ) was measured using the laser flash apparatus (Linseis) under a vacuum environment from 250K-900K. A thin disk (diameter 6 mm, thickness <0.7 mm) was cut from the cylindrical ingot to measure the thermal diffusivity in the same direction as that of the electrical conductivity and Seebeck coefficient. The *mass density* (ρ) was measured using the Archimedes method. The *thermal conductivity* (κ) was calculated using the relation $\kappa = \rho C_p \upsilon$. The electronic contribution to thermal conductivity was estimated from the electrical conductivity using the Wiedemann-Franz law.

**XRD Patterns and Lattice Parameters of MnTe-x%Cr**

XRD data is illustrated in Figure S1. XRD analysis shows the polycrystalline phase of MnTe-x%Cr with no phase impurity for x=3, and small traces of CrTe for the sample with x= 4, 5, 14, and 20. No evidence of Cr, MnO, or $MnTe_2$ phases is observed. The positive shifts of the diffraction lines towards a higher angle indicate Mn substitution by Cr in MnTe. The corresponding lattice parameters are extracted from the XRD data analysis using the whole powder profile fitting (WPPF) method with internal standard (for angular correction) and illustrated in Figure S1 as a function of x% of Cr in MnTe. The presence of CrTe phase and lattice parameter trends reveal that all Cr ions did not substitute Mn ions, and hence the actual substitution can be different from the nominal amount of Cr. Nonetheless, less substitution of Mn by Cr is not demerited the theoretical concept and experimental evidence for magnon-bipolar carrier drag effect. To avoid the confusion, we plotted the figures with nominal values, as the actual substitution is not known and not the focus of the study. As such, we avoid using the compositional formula like $Mn_{1-x}Cr_xTe$, which may raise the confusion of the complete substitution of Mn by Cr.

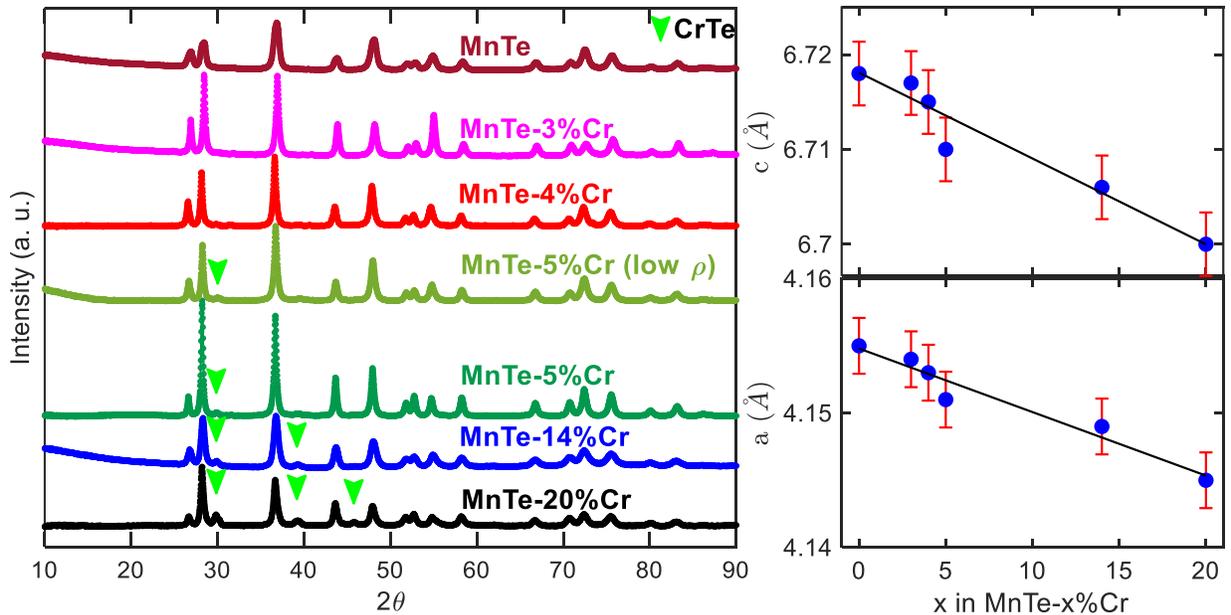

Figure S1: XRD diffraction pattern for MnTe-x%Cr samples (x = 0, 5, 4, 5, 14, and 20), and the extracted lattice parameters a and c as a function of Cr percentage. A 0.5% error bar is applied to the lattice parameters.

A comparison table is provided below to compare the obtained lattice parameters and magnetic transition temperature of MnTe-x%Cr samples studied in this work with previous literature.

Table S1: Comparison of the lattice parameters and magnetic transition temperatures of different Cr-doped MnTe samples with the previous data reported in the literature.

|  | Lattice Parameter, a (Å) / c (Å) | | | Transition Temperature, $T_C$ (K) | | | |
| --- | --- | --- | --- | --- | --- | --- | --- |
| x% Cr | This Work | Ref. 18 | Ref. 19 | This Work | Ref. 18 | Ref. 19 | Ref.24 |
| 3 | 4.154/6.718 | -/- | -/- | 174 | -- | -- | -- |
| 4 | 4.153/6.715 | -/- | 4.143/6.699 | 259 | -- | 305 | -- |
| 5 | 4.151/6.71 | -/- | -/- | 278 | -- | -- | ~180 |
| 8 | -/- | -/- | 4.139/6.673 | -- | -- | 305 | -- |
| 10 | -/- | 4.112/6.706 | -/- | -- | ~190 | -- | ~200 |
| 14 | 4.149/6.706 | -/- | 4.134/6.639 | 337 | -- | >380 | -- |
| 15 | -/- | -/- | -/- | -- | -- | -- | ~270 |
| 20 | 4.145/6.7 | 4.091/6.688 | -/- | 295 | ~210 | -- | -- |

**Optical Bandgap of MnTe-x%Cr samples**

According to the literature, Mn substitution by Cr dopant reduces the bandgap of the hexagonal MnTe phase [24,25]. Therefore, optical bandgap measurement can be one evidence for achieving the Mn substitution by Cr. Note that the actual substitution can be different from the nominal, as mentioned in the section for the XRD analysis. Figure S2 shows optical absorption spectra, $(\alpha h\upsilon)^2$, for direct transition as a function of photon energy, $h\upsilon$, for MnTe-3%Cr and MnTe-14%Cr samples along with pristine MnTe at room temperature. The bandgap was estimated from the intersect of the absorption spectrum with the zero-absorption line. The absorption spectrum is extrapolated to lower energy to get the intersection and estimate the bandgap, as shown in the figure. The plot shows a direct optical bandgap of approximately 1.25±0.12 eV for MnTe, which agrees with the literature [18,19], with a red-shift as the x increases. The optical bandgap sharply drops to 0.79eV with the addition of 3%Cr in MnTe. The red-shift of the bandgap due to the increment of Cr-doping has also been reported in the literature [18,24]. The estimated direct bandgaps for MnTe-x%Cr are shown in Figure S2, which are approximately 0.79±0.08 eV, and 0.7±0.07 eV at x = 3 and 14, respectively. All the values are calculated by a linear fit within the range shown in the figure. Cr can introduce an increment in the crystal-field splitting of the d-orbitals of MnTe, which can cause a decrease in the bandgap energy [18]. Also, the presence of some distinct absorption peaks in the UV-Vis spectra of MnTe-x%Cr samples, which are absent in the pristine MnTe, suggest the presence of possibly $3d^n$ ions like $Mn^{3+}$, $Cr^{3+}$, or $Cr^{2+}$ [42].

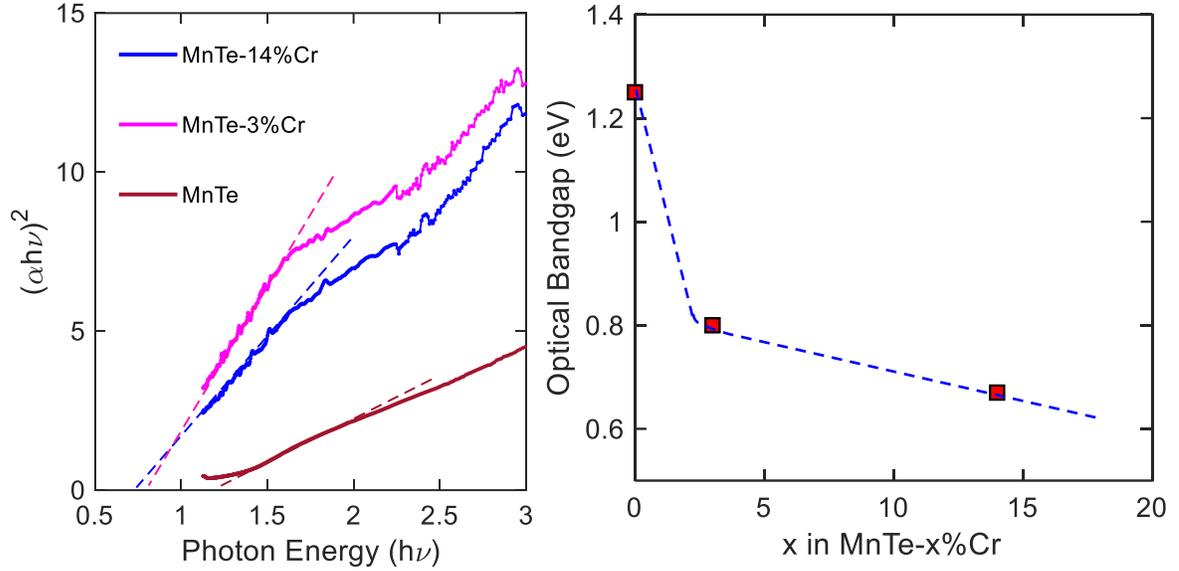

Figure S2: UV-Vis optical absorption spectrum for MnTe-x%Cr samples at room temperature.

**Hall Mobility Analysis for MnTe-x%Cr**

According to the carrier concentration data, low Cr-doped MnTe samples have bipolar carrier conduction, while 14% and 20% Cr-doped samples have dominant n-type and MnTe has dominant p-type carrier conduction. Using the single-carrier model, one can determine the Hall mobility for the majority carriers. As we do not have any information about the effective mass of the carriers, we utilized the single-carrier model for all the measured samples despite the bipolar nature of the low doped samples. Therefore, the information acquired about the carrier mobility for low Cr-doped MnTe samples can be misleading. Figure S3 demonstrates the calculated Hall mobility as a function of temperature and the logarithmic Hall coefficient as a function of the inverse temperature. According to [13], carrier mobility in MnTe is spin fluctuation dominated below the Neel temperature. At low temperatures, both Cr-doped MnTe samples, which have mixed contributions from electrons and holes, show lower mobility than undoped MnTe. Above 50K, the n-type carrier mobility of MnTe-20%Cr shows an ionized impurity scattering dominated trend, which saturates at around 1.2 cm$^2$/Vs, higher than the mobility of p-type MnTe at the same temperature. This indicates that electrons have higher mobility than holes in MnTe. This difference in the Hall mobility can also be understood from the logarithmic plot of the Hall coefficient as a function of inverse temperature [43]. The logarithmic value of the Hall coefficient, log($R_H$), gives comparatively higher value in n-type regime if $\mu_e/\mu_h \gg 1$, and vice versa [43]. In MnTe-4%Cr, the carrier mobility has mixed information from the mixed carrier transport, the modification in the intrinsic Mn-vacancy concentration in doped MnTe. According to the Hall coefficient plot, it is likely that both electrons and holes experience higher mobility in MnTe-4%Cr. However, the carrier mobility of low Cr-doped MnTe calculated from the single-carrier model can be different than the actual carrier mobility, which requires more detailed analysis beyond the scope of this work.

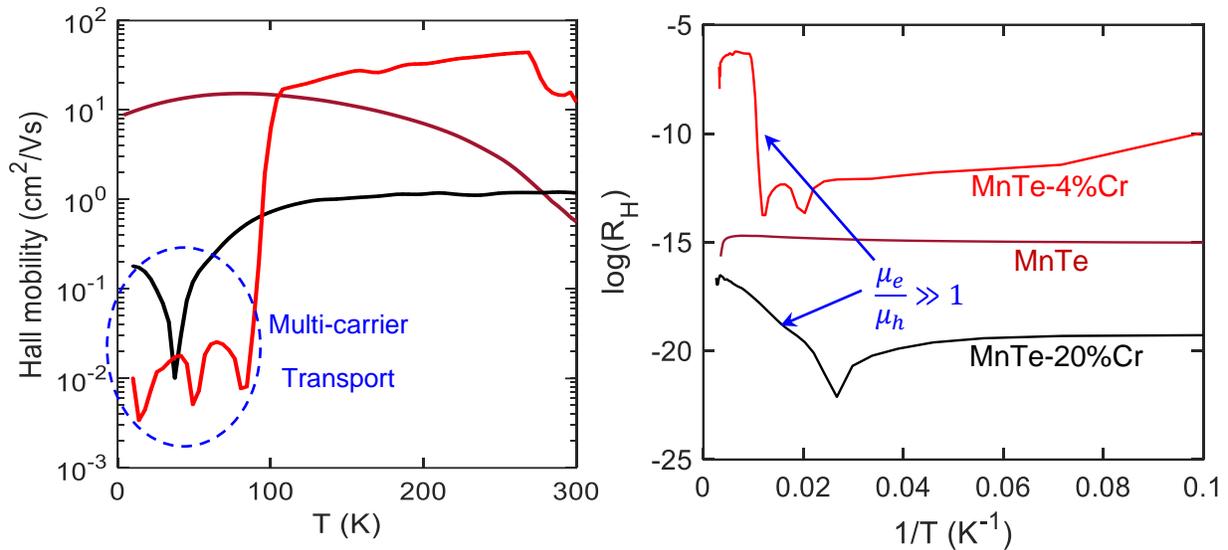

Figure S3: Temperature-dependent Hall mobility (left) and logarithmic Hall coefficient as a function of inverse temperature for MnTe-x%Cr samples (x = 0, 4, and 20).

**References**

[1] M. V. Costache, G. Bridoux, I. Neumann, and S. O. Valenzuela, Magnon-drag thermopile, Nat. Mater. 11, 199 (2012).
[2] S. J. Watzman, R. A. Duine, Y. Tserkovnyak, S. R. Boona, H. Jin, A. Prakash, Y. Zheng, and J. P. Heremans, Magnon-drag thermopower and Nernst coefficient in Fe, Co, and Ni, Phys. Rev. B 94, 144407, (2016).
[3] M. Sertkol, S. Ballıkaya, F. Aydoğdu, A. Güler, M. Özdemir, and Y. Öner, Thermoelectric and Magnetic Properties of Pt-Substituted $BaFe_{4-x}Pt_xSb_{12}$ Compounds, J. Electron. Mater. 46, 73 (2017).
[4] J. Heremans, Y. Zheng, H. Zhao, T. Lu, N. Liu, P. Sun, W. Wang, M. Rasoolianboroujeni, and D. Vashaee, High ZT in MnTe via spin physics, International and European Conference on Thermoelectric 2018, France, July 1-5, 2018.
[5] Y. Wang, N. S. Rogado, R. J. Cava, and N. P. Ong, Spin entropy as the likely source of enhanced thermopower in $Na_xCo_2O_4$, Nature 423, 425 (2003).
[6] G. Tang, F. Xu. Y. He, L. Wang, L. Qiu, and Z. Wang, Evidence for giant spin entropy contribution in thermoelectric $Ca_3Co_4O_{9+\delta}$, Phys. Stat. sol. (b) 250, 1327 (2012).
[7] A. G. Volkov, A. A. Povzner, V. V. Kryuk, and P. V. Bayankin, Spin fluctuations and properties of the thermoelectric power of nearly ferromagnetic iron monosilicide, Phys. Solid State 41, 960 (1999).
[8] R. Viennois, L. Girard, D. Ravot, S. Paschen, S. Charar, A. Mauger, P. Haen, and J. C. Tedenac, Thermoelectric properties of $Ce(La)Fe_4Sb_{12}$ skutterudites under a magnetic field, Phys. Rev. B 80, 155109 (2009).
[9] K. Vandaele, S. J. Watzman, B. Flebus, A. Prakash, Y. Zheng, S. R. Boona, and J. P. Heremans, Thermal spin transport and energy conversion, Mater. Today Phys. 1, 39 (2017).

[10] K. Sugihara, Magnon drag effect in magnetic semiconductors, J. Phys. Chem. Solids 33, 1365 (1972).
[11] B. H. Kim, J. S. Kim, T. H. Park, D. S. Lee, and Y. W. Park, Magnon drag effect as the dominant contribution to the thermopower in $Bi_{0.5-x}La_xSr_{0.5}MnO_3$ ($0.1 \leq x \leq 0.4$), J. Appl. Phys. 103, 113717 (2008).
[12] M. M. H. Polash, F. Mohaddes, M. Rasoulianboroujeni, and D. Vashaee, Magnon-drag thermopower in antiferromagnets versus ferromagnets, J. Mater. Chem. C, 8, 4049 (2020).
[13] Y. Zheng, T. Lu, Md M. H. Polash, M. Rasoulianboroujeni, N. Liu, M. E. Manley, Y. Deng, P. J. Sun, X. L. Chen, R. P. Hermann, D. Vashaee, J. P. Heremans, and H. Zhao, Paramagnon drag yields a high thermoelectric figure of merit in Li-doped MnTe, Sci. Adv. 5, eaat9461 (2019).
[14] D. Vashaee, M. M. H. Polash, V. Perelygin, M. Rasoulianboroujeni, Y. Zheng, T. Lu, N. Liu, M. Manley, R. Hermann, Alex I Smirnov, J. P. C. Heremans, and H. Zhao, Spin Effects Making zT > 1, Bulletin of the American Physical Society, Boston, Massachusetts, March 4–8, 2019.
[15] Y. Ren, J. Yang, Q. Jiang, D. Zhang, Z. Zhou, X. Li, J. Xin and X. He, Synergistic effect by Na doping and S substitution for high thermoelectric performance of p-type MnTe, J. Mater. Chem. C 5, 5076 (2017).
[16] Z. Zheng, X. Su, R. Deng, C. Stoumpos, H. Xie, W. Liu, Y. Yan, S. Hao, C. Uher, C. Wolverton, M. G. Kanatzidis, and X. Tang, Rhombohedral to Cubic Conversion of GeTe via MnTe Alloying Leads to Ultralow Thermal Conductivity, Electronic Band Convergence, and High Thermoelectric Performance, J. Am. Chem. Soc. 140, 2673 (2018).
[17] U. Sondermann, Magnetic and X-Ray Investigation in the system $V_xMn_{1-x}Te$ for x = 0 to x = 1, Z. Angew. Phys. 30, 41 (1970).
[18] G. I. Makovetski, A. I. Galyas, and K. I. Yanushkevich, Structural, magnetic, and electrical properties of solid solutions of the chromium telluride-manganese telluride system, Phys. Solid State 39, 280 (1997).
[19] Y. B. Li, Y. Q. Zhang, N. K. Sun, Q. Zhang, D. Li, J. Li, and Z. D. Zhang, Ferromagnetic semiconducting behavior of $Mn_{1-x}Cr_xTe$ compounds, Phys. Rev. B 72, 193308 (2005).
[20] F. K. Lotgering and E. W. Gorter, Solid solutions between ferromagnetic and antiferromagnetic compounds with NiAs structure, J. Phys. Chem. Solids 3, 238 (1957).
[21] N. H. Long and H. Akai, Ab-initio calculation of electronic and magnetic properties of $Mn_{1-x}Cr_xTe$, J. Supercond. Nov. Magn. 20, 473 (2007).
[22] M. Staruch and M Jain, Evidence of antiferromagnetic and ferromagnetic superexchange interactions in bulk $TbMn_{1-x}Cr_xO_3$, J. Phys.: Condens. Matter 26, 046005 (2014).
[23] Z. H. Wang, D. Y. Geng, W. J. Gong, J. Li, Y. B. Li, and Z. D. Zhang, Effect of adding Cr on magnetic properties and metallic behavior in MnTe film, Thin Solid Films 522, 175 (2012).
[24] K. J. Kim, J. H. Lee, H. J. Lee, J. Yoon, M.-H. Jung, W. Kim, S. J. Kim, and C. S. Kim, Effects of Cr Doping on the Electronic Structure of MnTe, J. Korean Phys. Soc. 53, 742745 (2008).
[25] H. Yadaka, T. Harada, and E. Hirahara, Electrical resistivity of chromium modified manganese telluride, J. Phys. Soc. Japan 17, 875 (1961).
[26] F. Qu, A. J. A. Beukman, S. N.-Perge, M. Wimmer, B.-M. Nguyen, W. Yi, J. Thorp, M. Sokolich, A. A. Kiselev, M. J. Manfra, C. M. Marcus, and L. P. Kouwenhoven, Electric and Magnetic Tuning Between the Trivial and Topological Phases in InAs/GaSb Double Quantum Wells, Phys. Rev. Lett. 115, 036803 (2015).
[27] M. H. Lee, S. Park, J. K. Lee, J. Chung, B. Ryu, S.-D. Park, and J.-S. Rhyee, Fine tuning of Fermi level by charged impurity-defect cluster formation and thermoelectric properties in n-type PbTe-based compounds, J. Mater. Chem. A 7, 16488 (2019).

[28] J. J. Gong, A. J. Hong, J. Shuai, L. Li, Z. B. Yan, Z. F. Ren, and J.-M. Liu, Investigation of the bipolar effect in the thermoelectric material $CaMg_2Bi_2$ using a first-principles study, Phys. Chem. Chem. Phys. 18, 16566 (2016).

[29] F. W. G. Rose, E. Billig, and J. E. Parrott, A Simple Derivation of the Thermoelectric Voltage in a Non-Degenerate Semiconductor, Int. J. Electron. 3, 481 (1957).

[30] A. Shahee, K. Singh, R. J. Choudhary, and N. P. Lalla, Evidence of ferromagnetic short-range correlations in cubic $La_{1-x}Sr_xMnO_{3-d}$ (x = 0.80, 0.85) above antiferromagnetic ordering, Phys. Stat. Sol. (b) 252, 1832 (2015).

[31] S. J. Kim, W. Kim, S. H. Bae, C. S. Kim, J. Yoon, M.-H. Jung, and K. J. Kim, Diluted Ferromagnetic Semiconductor in a Cr-Based MnTe Thin Film, J. Korean Phys. Soc. 52, 492 (2008).

[32] F. Gronvold, N. J. Kveseth, F. D. S. Marques, and J. Tichy, Thermophysical properties of manganese monotelluride from 298 to 700 K. Lattice constants, magnetic susceptibility, and antiferromagnetic transition, J. Chem. Thermodynamics 4, 795 (1972).

[33] N. E. Phillips, Low-temperature heat capacity of metals, Crit. Rev. Solid State Mater. Sci. 2, 467 (1971).

[34] S. Y. Dan'kov, A. M. Tishin, V. K. Pecharsky, and K. A. Gschneidner Jr., Magnetic phase transitions and the magnetothermal properties of gadolinium, Phys. Rev. B 57, 3478 (1998).

[35] C. Cervera, J. B. Rodriguez, J. P. Perez, H. Aït-Kaci, R. Chaghi, L. Konczewicz, S. Contreras, and P. Christol, Unambiguous determination of carrier concentration and mobility for InAs/GaSb superlattice photodiode optimization, J. Appl. Phys. 106, 033709 (2009).

[36] D. Mandrus, J. R. Thompson, R. Gaal, L. Forro, J. C. Bryan, B. C. Chakoumakos, L. M. Woods, B. C. Sales, R. S. Fishman, and V. Keppens, Continuous metal-insulator transition in the pyrochlore $Cd_2Os_2O_7$, Phys. Rev. B 63, 195104 (2001).

[37] C. Haas, Spin-Disorder Scattering and Magnetoresistance of Magnetic Semiconductors, Phys. Rev. 168, 531 (1968).

[38] E. D. Devyatkova, A. V. Golubkov, E. K. Kudinov, and I. A. Smirnov, The thermal conductivity of MnTe as affected by spin-phonon interaction, Soviet Phys. Solid State 6, 1813 (1964).

[39] S. Mu, R. P. Hermann, S. Gorsse, H. Zhao, M. E. Manley, R. S. Fishman, and L. Lindsay, Phonons, magnons, and lattice thermal transport in antiferromagnetic semiconductor MnTe, Phys. Rev. Mater. 3, 025403 (2019).

[40] M. M. H. Polash, V. Perelygin, M. Rasoulianboroujeni, Y. Zheng, T. Lu, N. Liu, M. Manley, R. Hermann, A. Smirnov, J. Heremans, H. Zhao, and D. Vashaee, Spin Effects Leading to zT>1: MnTe(Cr) vs MnTe(Li), 2019 MRS Spring Meeting, Phoenix, Arizona, April 22-26, 2019.

[41] F. J. Blatt, D. J. Flood, V. Rowe, P. A. Schroeder, and J. E. Cox, Magnon-drag Thermopower in Iron, Phys. Rev. Lett. 18, 395 (1967).

[42] T. Graf, M. Gjukic, M. S. Brandt, M. Stutzmann, and O. Ambacher, The $Mn^{3+/2+}$ acceptor level in group III nitrides, Appl. Phys. Lett. 81, 5159 (2002).

[43] C.-H. Ho, M.-H. Hsieh, and Y.-S. Huang, Compensation and Carrier Conduction in Synthetic $Fe_{1-x}Ni_xS2$ (0 ≤ x ≤ 0.1) Single Crystals, J. Electrochem. Soc. 155, H254 (2008).